\documentclass[twocolumn]{aastex63} %
\usepackage[latin1]{inputenc}
\usepackage{amsmath}
\usepackage{amsfonts}
\usepackage{amssymb}
\usepackage{subfigure}
\usepackage{graphicx}
\usepackage{xcolor}

\usepackage{natbib}
\newcommand{\steady}[0]{\textit{16TI} }
\newcommand{\spikes}[0]{\textit{40sp\_down} }

\listofchanges
\begin{document}

\title{Photospheric Prompt Emission From Long Gamma Ray Burst Simulations -- I. Optical Emission} 
\author{Tyler Parsotan$^1$ and Davide Lazzati}
\affiliation{Department of Physics, Oregon State University, 301 Weniger Hall, Corvallis, OR 97331, U.S.A.}

\begin{abstract} 
A complete understanding of Gamma Ray Bursts (GRBs) has been difficult to achieve due to our incomplete knowledge of the radiation mechanism that is responsible for producing the prompt emission. This emission, which is detected in the first tens of seconds of the GRB, is typically dominated by hard X-ray and gamma ray photons although, there have also been a few dozen prompt optical detections. These optical detections have the potential to discriminate between plausible prompt emission models, such as the photospheric and synchrotron shock models. In this work we use an improved MCRaT code, which includes cyclo-synchrotron emission and absorption, to conduct radiative transfer calculations from optical to gamma ray energies under the photospheric model. The calculations are conducted using a set of two dimensional relativistic hydrodynamic long GRB jet simulations, consisting of a constant and variable jet. 
{{
We predict the correlations between the optical and gamma ray light curves as functions of observer angle and jet variability and find that there should be extremely dim optical prompt precursors for large viewing angles. Additionally, the detected optical emission originates from dense regions of the outflow such as shock interfaces and the jet-cocoon interface. Our results also show that the photospheric model is not able to account for the current set of optical prompt detections that have been made and additional radiative mechanisms are needed to explain these prompt optical observations. These findings show the importance of conducting global radiative transfer simulations using hydrodynamically calculated jet structures.}} \newline 
\end{abstract} 


\section{Introduction}
Gamma Ray Bursts (GRBs), some of the most energetic events in the Universe \citep{Kulkarni_GRB_energy}, have been detected since the 1960's as pulses of gamma-ray radiation \citep{first_grbs}. These transient events have been classified into two groups based on duration times \citep{kouveliotou1993identification}; events that last $\lesssim 2$ s are known as Short GRBs (SGRBs) while events that last $\gtrsim 2$ s are classified as Long GRBs (LGRBs). SGRBs are associated with the merger of compact objects \citep{GW_NS_merger, grb_NS_merger_connection, lazzati2018_GRB170817_afterglow} and LGRBs are associated with core-collapse supernovae \citep{grb_sn_connection, grb_collapsar_model, hjorth2003_LGRB_SNe}. Both types of GRBs are thought to share a common physical mechanism that produces the prompt emission \citep{ghirlanda2011sgrb_lgrb_same_emission_mech}, the emission that is detected during the first few seconds of these events, however the details of the mechanism are still under investigation.

There are a number of models that attempt to describe the physical mechanism that produces the prompt emission. These can be broadly grouped into those that employ synchrotron radiation and those that rely on emission from the photospheres of the GRBs. The former set of models are based on and include the optically thin synchrotron shock model (SSM) \citep{SSM_REES_MES}, which describes shells of material that are launched by a central engine with varying speeds. These shells eventually collide with one another and produce nonthermal radiation if the optical depth is less than $1$. This model is able to account for the variability and nonthermal spectra observed in GRBs, however it in tension with the Yonetoku, Amati, and Golenetskii relationships \citep{Amati,Yonetoku,Golenetskii, ICMART_Zhang_2010}. In an attempt to remedy these discrepancies modifications have been made to the SSM through the use of globally ordered and random magnetic fields in GRB jets \citep{toma2008statistical_GRB_pol, ICMART_Zhang_2010}.

The other branch of models that is used to describe the prompt emission of GRBs relies on photospheric emission. This photospheric model considers photons produced deep in the GRB jet which interact with the matter in the jet; this radiation is eventually advected out of the jet once the optical depth is $\approx 1$ at the photosphere \citep{REES_MES_dissipative_photosphere, Peer_photospheric_non-thermal,Belo_collisional_photospheric_heating, lazzati_variable_photosphere}. Spectra calculated from the photospheric model can be non-thermal due to: subphotospheric dissipation and shocks \citep{Atul, parsotan_var, ito_mc_shocks}, photons originating from high latitude regions of the outflow \citep{parsotan_var, Peer_multicolor_bb}, and the photospheric region, where the photosphere is a volume of space (instead of a surface) where photons may experience their last scattering before being detected \citep{parsotan_mcrat, parsotan_var, Ito_3D_RHD, Peer_fuzzy_photosphere, Beloborodov_fuzzy_photosphere, ito2019photospheric}. The photospheric model is able to reproduce the Yonetoku, Amati, and Golenetskii relationships \citep{lazzati_photopshere, diego_lazzati_variable_grb, parsotan_mcrat, parsotan_var, ito2019photospheric} and many detected polarization measurements of GRBs \citep{parsotan_polarization}. Despite these successes, one drawback of the model has been its inability to consistently produce enough low energy photons (below typical GRB spectral peak energies of $\sim$ few $\times 10^2$ keV) in the outflow \citep{parsotan_mcrat, parsotan_var}. 

Theoretical studies of the photospheric model and the SSMs have primarily focused on the X-ray and gamma-ray energies in order to compare model predictions to observational data. As a result, various models are able to explain observed characteristics of GRBs in this energy regime. Alternatively, there have been a number of optical emission detections made during the prompt emission of GRBs which provide additional data for the purpose of model comparison. Optical data is less straightforward to interpret than gamma ray data due to the complications that it experiences from absorption along the observers line of sight and overlapping optical light emitted during the early afterglow phase of the GRB, due to forward/reverse shocks of the jet colliding with the external medium. 

\cite{kopac2013_prompt_optical} analyzed a sample of GRBs with optical detections during the prompt emission and found a diverse set of observed properties. The varied behavior that they found can be attributed to the various physical mechanisms that produced the optical emission of the GRBs in their sample. Some of the optical emission in their sample is attributed to the prompt emission, due to the optical and X-ray emissions tracking one another, while other detections can be attributed to emission from the early afterglow phase of the GRB, due to similarities with emission expected from forward or reverse shocks \citep{Sari_1999_afterglow_reverse_shock, meszaros1999_afterglow_reverse_shock}. \cite{Oganesyan2019_prompt_opt} recently refined this analysis and only analyzed GRBs with optical emission that was thought of as being directly related to the GRB prompt emission. {{In the spirit of having a concise list of GRBs with optical prompt emission detections that have been deemed to be produced by the same emission mechanism responsible for the gamma ray prompt emission}}, we have expanded on the work of \cite{kopac2013_prompt_optical} and \cite{Oganesyan2019_prompt_opt} by identifying GRBs in the literature that have optical detections that have been attributed to the prompt emission, including those analyzed by \cite{Oganesyan2019_prompt_opt}; this information is compiled into Table \ref{opt_reference_table} where we list the name of the GRB, the instrument that first took optical observations, the redshift of the GRB where available, the peak observed magnitude of the observed optical emission (independent of filter), {{the spectral index, $\beta_1$, measured from the optical to X-ray bands where available or where an estimate could be made}}, and the appropriate reference.

\begin{deluxetable*}{lccccl}
\tablecaption{{Various optical detections of GRBs that have been associated with the emission mechanism that produces gamma-ray prompt emission in the literature}. We list the GRB, the instrument that first started optical observations, the redshift of the GRB where available, the peak observed Vega magnitude of the observed optical emission, {{the spectral index, $\beta_1$, from the optical to X-ray energies in the spectrum (such that $f_\nu \propto \nu^{\beta_1}$) where available}} , and the appropriate reference for the GRB. {The $\beta_1$ obtained below from the cited works were either reported in the manuscript or estimated by us based on plots of the prompt spectra; in the cases of GRBs 041219A and 050401, we calculated the spectral slope from \citeauthor{rykoff2005_grb050401}'s (\citeyear{rykoff2005_grb050401}) Table 3. When various time resolved spectra that extended to optical energies are presented, we estimate the $\beta_1$ and present them as ranges, such as in the case of GRB 080319B. In some cases, such as for GRB 080928, the low energy tails of the spectra were fixed to be that of synchrotron radiation in order to acquire a converged spectral slope power law fit. Due to the risk of oversimplifying these spectral fits, we have excluded these spectral indexes from the Table.} \label{opt_reference_table}}
\tablenum{1}
\tablehead{\colhead{GRB} & \colhead{Instrument} & \colhead{z} & \colhead{Peak Magnitude (Vega)} & \colhead{$\beta_1$} & \colhead{Reference}}
\startdata
041219A & RAPTOR\tablenotemark{a} & 0.31 & 13.7 & $\sim -0.03$ & \cite{vestrand2005_grb041219A} \\
050319 & SWIFT UVOT\tablenotemark{b} & 3.24 & 17.5 & $-0.8$ & \cite{cusumano2006swift}  \\
050401 & ROTSE-IIIa\tablenotemark{c} & 2.9 & 16.8 & $\sim -0.17$ & \cite{rykoff2005_grb050401} \\
050820A & RAPTOR & 2.615 & 14.6 & $-0.96$ - $-.25$ &\cite{vestrand2006_grb050820A} \\
051111 & KAIT\tablenotemark{d} & 1.55 & 13.6 & - &\cite{butler2006_grb051111} \\
060526 & Watcher Telescope\tablenotemark{e} & 3.2 & 16.87 & - &\cite{thone2010_grb060526}  \\
060814 & TAROT\tablenotemark{f} & - & $>14.8$ & - &\cite{Oganesyan2019_prompt_opt} \\
061121 & Swift UVOT & 1.314 & 15.8 & $-0.11$ &\cite{page2007_grb061121} \\
080310 & Swift UVOT & 2.43 & 17.1 & $-0.65$ - $1$ &\cite{littlejohns2012_grb080310} \\
080319B & ``Pi of the Sky''\tablenotemark{g} & 0.937 & 5.3 & $-3.6$ - $-2.7$ &\cite{Racusin2008_grb080319B} \\
080905B & Watcher & 2.374 & $\sim 15$ & - &\cite{ferrero2010_watcher_grb080905B} \\
080928 & Swift UVOT & 1.692 & 18.44 & - &\cite{rossi2011_grb080928} \\
090715B & Swift UVOT & - & 20.37 & - &\cite{Oganesyan2019_prompt_opt} \\
090727 & Liverpool Telescope\tablenotemark{h} & - & 18.89 &$\sim -1.31$ - $-0.29$ &\cite{kopac2013_prompt_optical} \\
100901A & MASTER\tablenotemark{i} & 1.408 & 17 & - &\cite{gorbovskoy2012_grb100901A} \\
110102A & Swift UVOT & $<2.5$ & 17.6 & - &\cite{Oganesyan2019_prompt_opt} \\
110119A & Swift UVOT & - & 16.5 & - &\cite{Oganesyan2019_prompt_opt} \\
110205A & Swift UVOT & 2.22 & 17.47 & 0.13 &\cite{cucchiara2011_grb110205A} \\
111103B & Swift UVOT & - & $> 19.1$ & - &\cite{Oganesyan2019_prompt_opt} \\
111123A & Swift UVOT & - & $> 20.1$ & - &\cite{Oganesyan2019_prompt_opt} \\
121123A & Swift UVOT & $1.5-3.4$ & 18.58 & - &\cite{Oganesyan2019_prompt_opt} \\
130514A & GROND\tablenotemark{j} & 3.6 & 14.96 & - &\cite{Oganesyan2019_prompt_opt} \\
140108A & Swift UVOT & - &20.17 & - &\cite{Oganesyan2019_prompt_opt} \\
140206A & Swift UVOT & $\sim 2-3$ & 15.88 & - &\cite{Oganesyan2019_prompt_opt} \\
140512A & TNT\tablenotemark{k} & 0.725 & 13.09 & $\sim 0.32$ &\cite{huang2016_grb140512A} \\
151021A & Skynet\tablenotemark{l} & - & $>15.7$ & - &\cite{Oganesyan2019_prompt_opt} \\
160625B & MASTER & 1.406 & 7.9 & $-0.52$ &\cite{troja2017_grb160625B} \\
\enddata
$^\mathrm{a}$\cite{vestrand2002raptor} $^\mathrm{b}$\cite{roming2005swiftuvot} $^\mathrm{c}$\cite{akerlof2003rotse} $^\mathrm{d}$\cite{Filippenko_KAIT} $^\mathrm{e}$\cite{Ferrero_2010_watcher} $^\mathrm{f}$\cite{klotz2009_tarot} $^\mathrm{g}$\cite{cwiok2007_piofthesky} $^\mathrm{h}$\cite{steele2004liverpool} $^\mathrm{i}$\cite{lipunov2010master} $^\mathrm{j}$\cite{greiner2008grond} $^\mathrm{k}$\cite{zheng2008tnt} $^\mathrm{l}$\cite{smith2016skynet}

\end{deluxetable*}

The models that have attempted to describe GRB optical prompt emission employ synchrotron radiation mechanisms (see e.g. \cite{ICMART_Zhang_2010} and \cite{zhao2009_inversecompton_opticalprompt}); although, \cite{zheng2006comptonization_promptoptical} did propose that the optical prompt emission from GRB 041219A was formed by saturated comptonization of the GRB's gamma ray photons in a surrounding electron cloud. As a result, most of the analysis that has been conducted of prompt optical emission has been under the assumption of synchrotron emission producing the detected radiation, including the analysis conducted by \cite{kopac2013_prompt_optical} and \cite{Oganesyan2019_prompt_opt}. In their analysis, \cite{Oganesyan2019_prompt_opt} fit the spectra of the GRBs in their sample set with two models: a synchrotron model, that captured the physics of synchrotron radiation for a range of values of interest, and a ``thermalized'' two component model, consisting of a cutoff power law with a subdominant blackbody component. They found that the synchrotron models provided superior fits to the spectra from optical to gamma rays, while the ``thermalized'' model typically overpredicted the optical portion of the spectra. When using the synchrotron model to predict the properties of the emission region of the jet (such as bulk lorentz factor, number density of electrons, magnetic field strength, etc.), \cite{Oganesyan2019_prompt_opt} calculated values that were in tension with other analysis of GRB prompt emission weakening their prediction that synchrotron emission is responsible for the prompt emission of GRBs. Furthermore, \cite{Oganesyan2019_prompt_opt} point out that relaxing the synchrotron emission mechanism criteria and permitting synchrotron self compton solutions would produce optical prompt emission that is brighter than what is typically observed. {{Synchrotron emission is still commonly taken to be the leading model for explaining optical prompt emission, especially due to the brightness of the naked-eye burst GRB 080319B \citep{Racusin2008_grb080319B, shen_synch_optical_emission, Tang_2006_optical_lag}. There are many variations of the synchrotron emission mechanism that have been invoked to explain observed GRB optical prompt emission. \cite{kumar_relativistic_turb_grb080319B} invoke relativistic turbulence and synchrotron emission due to the turbulence to explain the optical emission observed in GRB 080319B while \cite{Fan_2009_neutron_proton_shells} used synchrotron emission from proton/neutron shells colliding in the outflow to explain the optical prompt emission. \cite{Li_Waxman_residual_collisions_optical} theorized that synchrotron emission produced by residual internal shocks would also be able to account for observed optical prompt emission. Optical prompt emission has also been interpreted as synchrotron emission that has been produced in regions of the jet that are distinct from the region where gamma ray photons are produced \citep{Fan_2009_neutron_proton_shells, Zou_2009_optical_different_region}. Inverse compton has been proposed as a plausible radiation mechanism by \cite{Racusin2008_grb080319B} however this model has been shown to violate energy constraints obtained from GRB observations \citep{piran_IC_energy_limits}. While synchrotron emission is still the leading emission mechanism for explaining the current set of optical prompt emission detections, there has not been sufficient investigation of the photospheric model and its ability to account for the few dozen optical prompt detections. }}

The photospheric model is in a prime position for making predictions of the optical prompt emission. Although this model had been neglected in the past, {recent} advances in making predictions under this model has made the task of simulating emission across the electromagnetic spectrum feasible. Current state-of-the-art radiative transfer calculations of photospheric prompt emission combines complex jet structures and self-consistent radiation treatments \citep{MCRaT, parsotan_mcrat, parsotan_var, Ito_3D_RHD, ito2019photospheric, parsotan_polarization} providing a deeper insight into the emission from GRBs. The addition of cyclo-synchrotron emission and absorption processes into these radiative transfer calculations, as presented in this paper, make it possible to {{explore how much optical emission is expected under the photospheric model and}} make predictions of the optical prompt emission under this model.

In this paper we present the first simulations of the photospheric optical prompt emission for a set of LGRB special relativistic hydrodynamic (SRHD) simulations. We show synthetic spectra and light curves for these GRBs and the varied degree of correlation between the optical light curves and the bolometric light curves, which are dominated by gamma ray radiation. Additionally, we explore where the optical photons originate in the outflow. In Section \ref{global_methods}, we discuss the radiative transfer simulations, the mock observations, and the hydrodynamic simulations that the radiative transfer calculations used to define the jet profile. In Section \ref{results} and \ref{summary}, we outline the results of our radiative transfer simulations and their implications on our understanding of current and future optical prompt emission measurements, respectively. In a companion paper, we extend this analysis to polarization and conduct a time resolved spectro-polarimetry analysis of GRB prompt emission from optical to gamma rays \citep{Parsotan_spectropolarimetry}.

\section{Methods}
\label{global_methods}

\subsection{The MCRaT Code}
The MCRaT (Monte Carlo Radiation Transfer) code{\footnote{The MCRaT code is open-source and is available to download at: https://github.com/lazzati-astro/MCRaT/}}{{\citep{parsotan_mcrat_software_2021_4924630}}} is an open source, highly parallelized radiative transfer code that conducts post-processing  radiative transfer calculations using a jet profile acquired from special relativistic hydrodynamic (SRHD) simulations of GRB jets \citep{parsotan_mcrat, parsotan_var, parsotan_polarization}. The code is compatible with jets simulated with the FLASH and PLUTO hydrodynamics codes \citep{fryxell2000flash, pluto_amr} and it takes into account a number of physical processes relevant to jetted systems such as Compton scattering, the Klein-Nishina cross section with polarization, and cyclo-synchrotron emission and absorption (as described in Section \ref{cyclo-synch}; \cite{MCRaT,parsotan_mcrat, parsotan_var, parsotan_polarization}). 

The MCRaT code injects a  blackbody or Wien spectrum into the SRHD simulation where the optical depth is $\gtrsim 10^3$ or $\gtrsim 10^2$ respectively \citep{parsotan_var}. The injected photons are then individually scattered and propagated from frame to frame of the SRHD simulation. The properties of the jet are acquired from the SRHD simulation and are used to determine: which photons will scatter, the energy of the electron that the photon will scatter with, and the lab frame energies of the photons once the scattering is complete.

\subsection{Cyclo-synchrotron Emission in MCRaT} \label{cyclo-synch}
\begin{figure*}[]
 \centering
 \fig{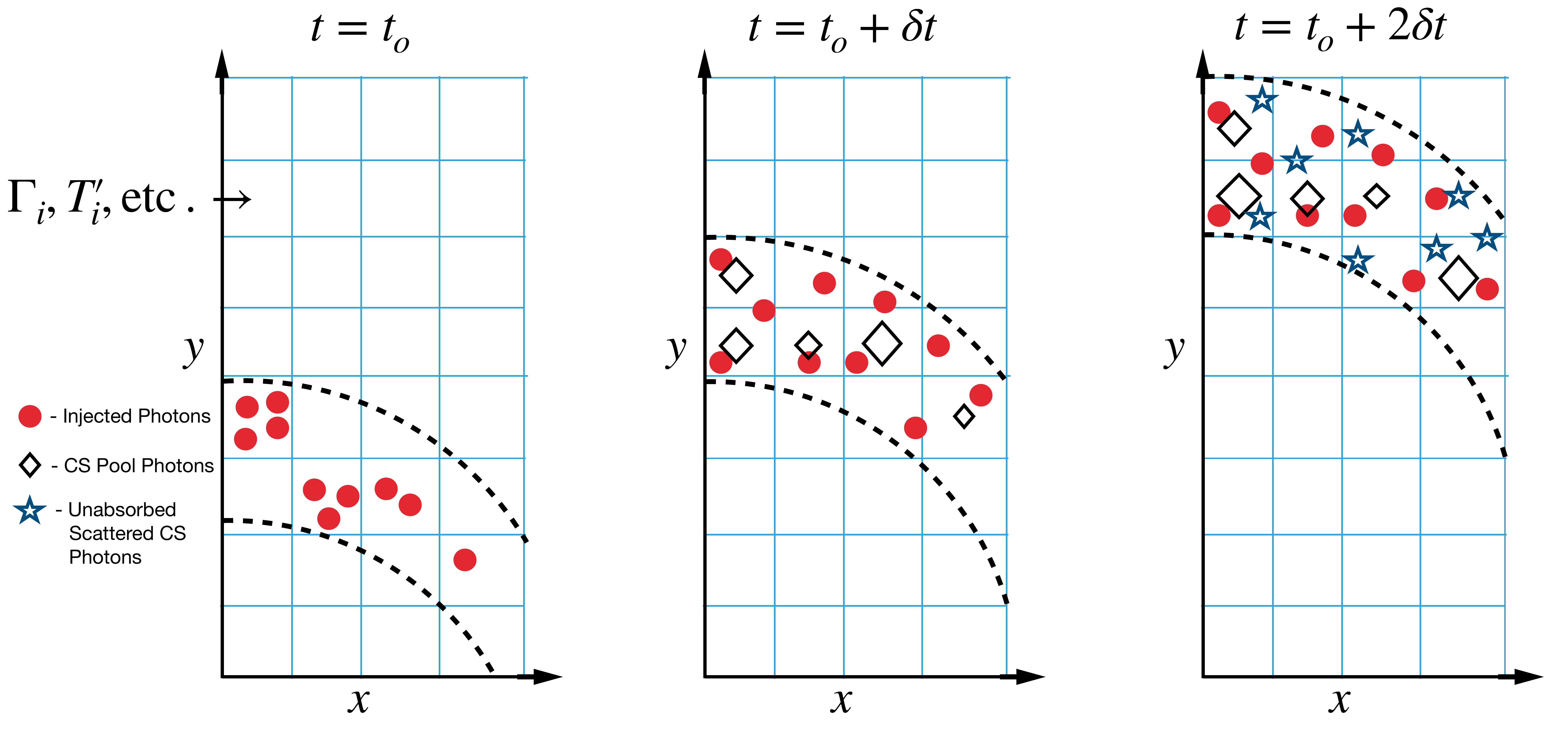}{\textwidth}{}
 \caption{A cartoon showing the progression of a typical MCRaT simulation with CS emission and absorption. We show the SRHD simulation fluid elements where various fluid parameters are defined, the thermally injected photons as red circles, the CS pool of photons as black diamonds with the size of the diamond correponding to the number of pool photons located within a cell, and scattered, unabsorbed photons as blue stars. We also show the shell that defines where the pool of CS photons are emitted into the simulation as dashed black lines. In the first frame of the simulation, at $t=t_o$, the code injects only thermal photons with the number of photons dependent on the temperature of the cell. In the next frame, at $t=t_o+\delta t$, the code emits the pool of CS photons and scatters them appropriately. At $t=t_o+2\delta t$ the pool of CS photons are absorbed along with any other photons with energies less than the CS frequency of the cell that they are located within. Then, the next frame is loaded and a new set of CS pool photons are emitted to be potentially scattered alongside the original thermally injected photons and the scattered CS photons that have survived absorption.} 
 \label{photon_population_fig}
\end{figure*}
We have enhanced the MCRaT code by including cyclo-synchrotron (CS) emission and absorption, allowing us to simulate the prompt emission of GRBs from optical to gamma ray energies. In this section we will briefly describe our implementation of these processes in the code and show tests of the algorithm. Details of the implementation of CS emission and absorption can be found in Appendix \ref{cs_appendix}.

From frame to frame of the SRHD simulation, the photons propagate a distance $c\delta t$, where c is the speed of light and $\delta t$ is the time between two simulation frames. This defines a shell where we emit a pool of CS photons into the MCRaT simulation\footnote{The emission of CS photons does not occur in the very first frame of the SRHD simulation due to the fact that we typically inject photons drawn from a blackbody spectrum as our initial condition, which accounts for all emission and absorption processes.}. We identify the fluid elements within this shell and calculate the number of CS photons that will be emitted into each element. Within a given fluid element, we calculate the number of CS photons that will be emitted by integrating the blackbody number density spectrum up to the cyclotron frequency of that given cell. {{Since we inject CS photons based on the local blackbody spectrum, we are naturally limited by the number of photons that would be expected under the blackbody distribution and can never exceed that flux.}} The comoving magnetic field of the fluid element, which affects the cyclotron frequency of the same element, can be calculated in two ways: 1) using the internal energy of the fluid element, or 2) using the total energy of the fluid parcel. As outlined in Appendix \ref{cs_appendix}, the fluid frame magnetic field in the internal energy case is calculated by setting the comoving magnetic field energy density of the fluid element equal to the electron thermal kinetic energy of that same element, with some unitless factor $\epsilon_{B}$. This results in
\begin{equation}
B_i'^2=8\pi\frac{3}{2}n_ikT_i'\epsilon_{B}
\end{equation}
where $n_i$ is the number density of electrons in the fluid element, and the comoving temperature of the fluid element $T_i'=(3p_i/a)^\frac{1}{4}$, where $p_i$ is the pressure of the fluid element as acquired from the SRHD simulation and $a$ is the radiation density constant. In the total energy case the comoving magnetic field is calculated by relating the luminosity of the magnetic field  to the total luminosity of the fluid element with a unitless parameter $\epsilon_{B}$. Here, the comoving magnetic field is given as
\begin{equation}
B_i'^2=8\pi(\rho_i c^2+4p_i)\epsilon_{B} 
\end{equation}
where $\rho_i$ is the density of the fluid element.

With the number of CS photons calculated to be in each cell of the shell completed, the total number of photons that should be emitted into the shell can be calculated as the sum of the CS photons emitted into each fluid element. The code then follows a procedure that attempts to balance the actual calculated number of photons that should be emitted into the shell with a computationally feasible number of MCRaT photons. This procedure iteratively calculates the number of photons in each cell while varying the weight of the photons (the weight is the number of physical photons each MCRaT photon packet represents) until the total number of photons emitted into the shell falls within a user defined range.

The aforementioned procedure emits a pool of CS photons into the MCRaT simulation, which is a stagnant population of CS photons that can potentially be scattered as the simulation continues. There are two other populations of photons in MCRaT: the thermally injected blackbody photons that make up the initial condition of the simulation and the scattered CS photons, which may have energies that are larger than the cyclotron frequencies of the fluid elements that they are located in and thus will survive absorption.

After creating the pool of CS photons, MCRaT continues its algorithm of: calculating the mean free path of photons, incrementing the distance traveled of each photon that is not part of the CS pool to correspond to the smallest mean free path, and scattering the photon with this mean free path. This cycle is continued until the simulation time has been incremented to correspond to that of the next SRHD simulation frame. If a pool CS photon has been chosen to be scattered, it will be replaced in the pool of CS photons. Then the photon will be randomly placed within its fluid element and allowed to scatter normally. Once a time $\delta t$ has passed in the simulation, the code checks to see if any photons have frequencies that are smaller than the cyclotron frequency of the fluid element that they are located within at that time. If this condition is true, those photons are ``absorbed'' by removing them from the list of photons that are considered for scattering in MCRaT. 

The various photon populations that exist in MCRaT and their behaviors as the simulation progresses are outlined in Figure \ref{photon_population_fig}. We show the initial thermally injected blackbody photons as red circles, the pool of CS photons as black diamonds, and the scattered CS photons that survive absorption as blue stars. We also plot the shell that marks where the CS pool of photons are emitted into the simulation as the black dashed lines. In this cartoon, we show the injection of thermal photons into the first frame of the simulation, at $t=t_o$, based on the properties of each fluid element, where the hottest and coolest fluid elements corresponds to high and low photon densities respectively \citep{parsotan_var}. In the next frame of the SRHD simulation, at $t=t_o+\delta t$, the original set of thermal photons still exist but there is also the pool of CS photons that can potentially be scattered to higher energies. Here the size of each diamond marker represents the amount of CS pool photons that are present in each fluid element, which is dependent on $B_i'$. Just before the next frame is loaded, corresponding to $t=t_o+2\delta t$, MCRaT ``absorbs'' the CS pool photons and any other photons that have energies smaller than the CS frequency of the fluid element that they are located within. After, the next frame of the simulation is loaded and the original thermally injected photons are present and so are some scattered CS photons that have survived absorption. In this new frame, the shell denoting where the CS photons are emitted has moved and we have placed a new set of pool CS photons that can potentially be upscattered. 

As the simulation progresses, the number of photons will naturally grow. In order to keep the number of simulated photons reasonable, MCRaT periodically rebins the scattered CS photons in energy and space (as detailed in Appendix \ref{rebinning}).

In order to verify the proper implementation of the CS process in MCRaT and to understand the effects that the various magnetic field calculations (the internal energy and total energy options as detailed in Section \ref{calc_b}) have on the spectra, we ran a number of test simulations of MCRaT on a spherical outflow, similar to the analysis done by \cite{MCRaT}. These tests and their results are outlined in Section \ref{verify_cs}, which we summarize briefly. We find that the total energy method of calculating the magnetic field emits a larger number of CS photons in the MCRaT simulation compared to the internal energy method. We also find that our MCRaT spectra are able to recover the saturated comptonized spectra for an input of soft photons, which was also recovered by \cite{vurm2013thermalization}. Additionally, we find that the fraction of energy in the magnetic field compared to the total energy of the outflow, $\epsilon_{B}$, does not have a large effect on the resultant spectrum; this result means that we can set $\epsilon_{B}=0.5$, as we do in this paper, and not expect our results to change much for $\epsilon_{B}$ that are a factor of $\sim 2$ different from $0.5$.

\subsection{Mock Observations}
The results of the MCRaT simulations can be used to produce mock observations of the SRHD LGRB simulations. The procedure for conducting these mock observations has been outlined in \cite{parsotan_mcrat}, \cite{parsotan_var}, and \cite{parsotan_polarization}. Here, we briefly summarize the methods we use to construct our mock observations, which is encoded in our ProcessMCRaT code\footnote{The code used to conduct the mock observations is also open source and is available at: https://github.com/parsotat/ProcessMCRaT} {{\citep{parsotan_processmcrat_software_2021_4918108}}}.

Mock observed light curves are produced by calculating the time of arrival of each MCRaT photon with respect to a virtual detector located at some distance from the origin of the SRHD simulation. We place the virtual detector at a number of viewing angles, $\theta_\mathrm{v}$, with respect to the jet axis and calculate which photons are propagating towards an observer located at that angle. We accept photons that are moving towards the observer within a $\pm 0.5^\circ$ acceptance interval. The light curves can be binned into uniform time bins or into variable time bin sizes that are determined by a bayesian blocks binning algorithm \citep{astropy:2013, astropy:2018}. We present bolometric and optical light curves in this paper, each being binned into the same time bins as the other. Bolometric light curves are calculated following the aforementioned method and no photon energies are excluded from the calculation; the bolometric light curve is typically dominated by the gamma ray emission. The optical light curves that we show in this paper are calculated in the same manner except we only consider photons that have energies that fall within the Swift UVOT White bandpass, from 1597-7820 $\AA$\footnote{http://svo2.cab.inta-csic.es/theory/fps/} ($\sim 1.5-7.7$ eV) \citep{poole2008_swiftphotometric,rodrigo2020svo}. This choice is motivated by Table \ref{opt_reference_table}, where we find that Swift UVOT has measured the majority of the optical prompt emission detections. 

Time resolved or time integrated spectra, in units of counts, are produced by binning photons based on their lab frame energies. These spectra are then fitted with either a Comptonized (``COMP'') function{\footnote{{While this function is typically called the Comptonized function in observational studies, the function is better described as a cutoff powerlaw. Comptonization can produce many types of spectra and cutoff power laws can be produced for a number of emission processes.}}} \citep{FERMI} or a Band function \citep{Band}; we fit the energy bins with $\ge 10$ photons in order to approximate errors in the bins as being gaussian. Unlike prior studies where the entirety of the spectrum is fit \citep{parsotan_mcrat, parsotan_var}, we fit the spectra that we acquire from 8 keV to 40 MeV which is typically done in observational analysis of GRB spectra \citep{FERMI}. 

The mock observations can also be related to the GRB jet structure as acquired by the SRHD GRB simulation. We do this by relating the time of the mock observation of interest to the equal arrival time surfaces (EATS) of the SRHD simulated jet. The EATS are acquired by calculating the location that photons would be emitting along a given observers line of sight for the time of interest in the light curve \citep{parsotan_polarization}. 

{{An important assumption that goes into constructing the mock observations is that the MCRaT photons are decoupled from the fluid in the jet by the end of the MCRaT simulation, which is not always the case \citep{parsotan_mcrat, parsotan_var, parsotan_polarization}. The only effect that this has on the mock observables is that the photons have not fully cooled, meaning that they are more energetic than they should be \citep{parsotan_var}.}}

\subsection{The Simulation Set}
The FLASH 2D SRHD simulations set analyzed in this work are identical to the ones used by \cite{parsotan_polarization}. Here, we briefly summarize these SRHD simulations and the MCRaT simulations that were conducted for this paper.

We ran MCRaT on two SRHD simulations where a jet was launched into a 16TI progenitor star with a density profile given by \cite{Woosley_Heger}. The first simulation, which we name \steady, has a domain of $2.5 \times 10^{13}$ cm along the direction of the jet. The \steady simulation jet was injected with a constant luminosity, $L_\mathrm{inj}=5.33 \times 10^{50}$ erg/s, for 100 s at a radius of $1\times 10^9$ cm, with an initial lorentz factor of 5, an opening angle of $10^\circ$, and an internal over rest-mass energy ratio, $\eta=80$ \citep{lazzati_photopshere}. The other simulation, which we denote as \spikes, has a jet that is injected with similar parameters as the \steady simulation with the exception of the time that the jet was on and the temporal structure of the jet. In this case, the jet was on for a total of 40 s and the injected jet profile consisted of 40 half second pulses each followed by another half second of quiescence. The luminosity of each active pulse of the jet is decreased by 5\% with respect to the initial pulse \citep{diego_lazzati_variable_grb}. The domain of the \spikes simulation is a factor of 10 smaller than the domain of the \steady simulation. After 40 s and 100 s in the \spikes and \steady simulations respectively, the jets were turned off and allowed to evolve for a few hundred more seconds.

The MCRaT calculations were conducted for the period of time that the jet was on in each SRHD simulation. We set MCRaT to calculate the magnetic field for CS emission and absorption using the total energy option, in which we set $\epsilon_{B}=0.5$ for all the calculations presented in Section \ref{results}. The initial injected spectrum into the MCRaT simulations was set to be a blackbody spectrum. Since the number of photons in the MCRaT simulations can change due to the emission, absorption, and rebinning processes in the code, we report the number of photons that exist at the end of the given MCRaT simulation. The \steady simulation ended with a total of  $\sim 10^7$ photons while the \spikes simulation ended with a total of $\sim 6\times 10^6$ photons. The average number of scatterings that photons experienced by the end of each simulation was $\sim 4\times 10^3$ for the \steady simulation and $\sim 2\times 10^5$ for the \spikes simulation, showing that the optical depth of where we started our MCRaT simulation in the SRHD jets are appropriate for a blackbody spectrum \citep{parsotan_var}. We produced mock observed light curves and mock observed time integrated and time resolved spectra for observers located at a range of $\theta_\mathrm{v}$, from $1-15^\circ$ for the \steady simulation and from $1-9^\circ$ for the \spikes simulation.

\section{Results} \label{results}
In this section, we will outline some of the results that we have obtained regarding optical prompt emission from our MCRaT calculations. We will first present MCRaT calculated light curves and spectra and then delve into the origins of the detected photons in the simulated GRB jets.

\subsection{Bolometric and Optical Light Curves} \label{section_lc}
The bolometric light curves that we calculate from our MCRaT simulations are similar to the light curves acquired in previous works \citep{parsotan_mcrat, parsotan_var}. Here, we show the relationship that we find between the bolometric and optical light curves for the \steady and \spikes simulations which we acquire by calculating various time signal properties such as correlation and time lag between each light curve.

\begin{figure*}[]
 \centering
 \gridline{
 \fig{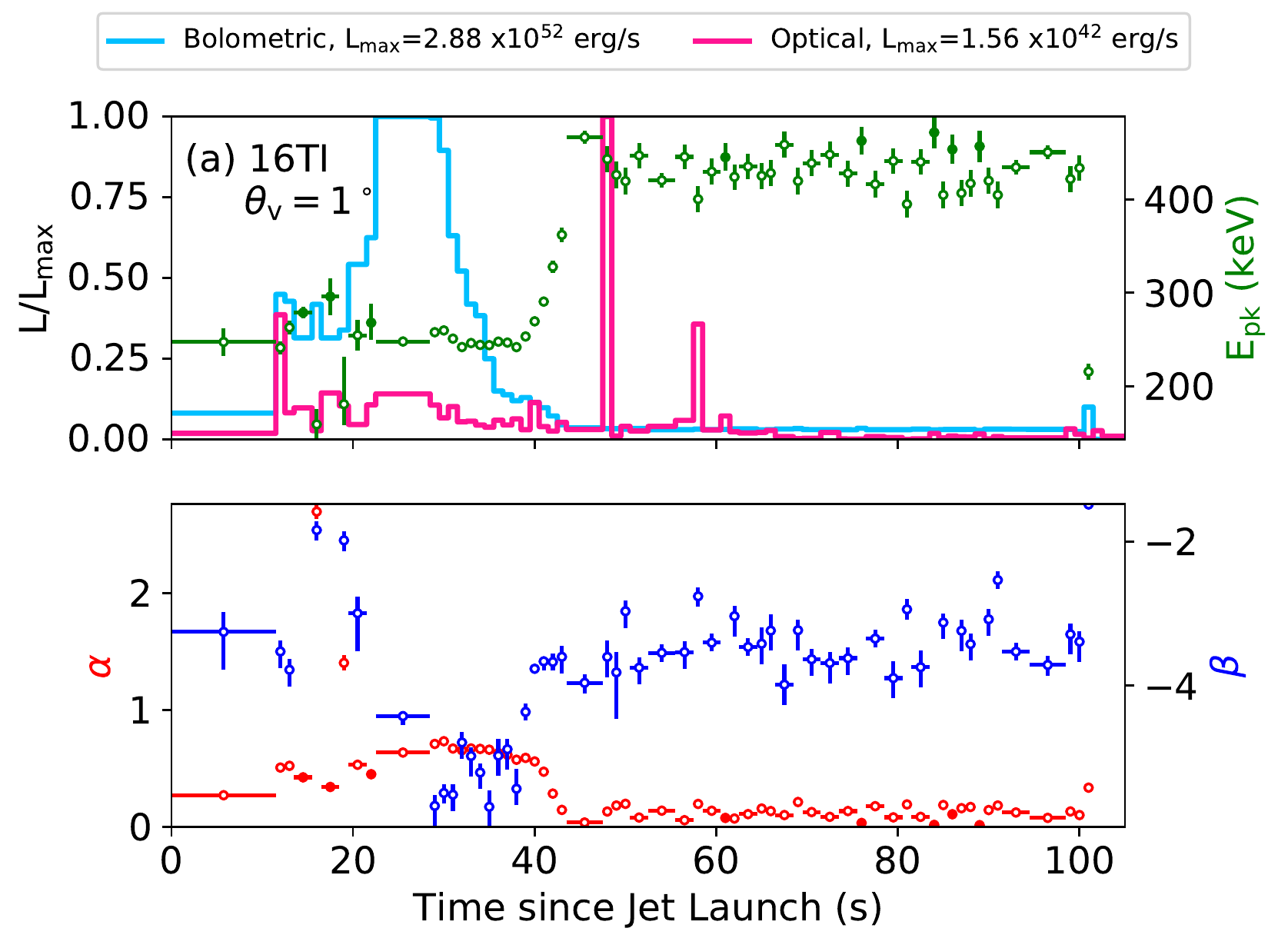}{0.5\textwidth}{\label{16ti_1_lc}}
 \fig{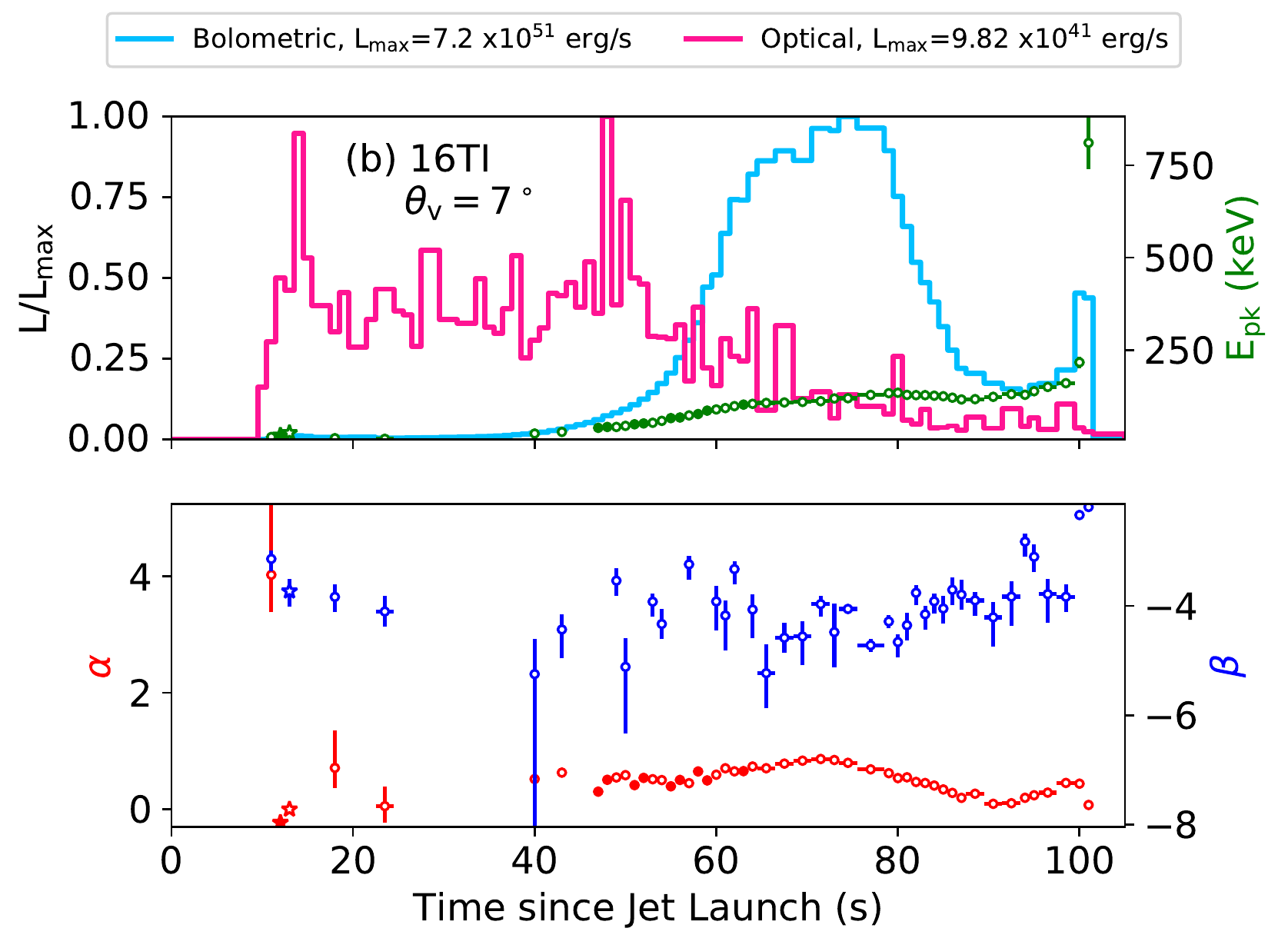}{0.5\textwidth}{\label{16ti_7_lc}}
 }
 \gridline{
  \fig{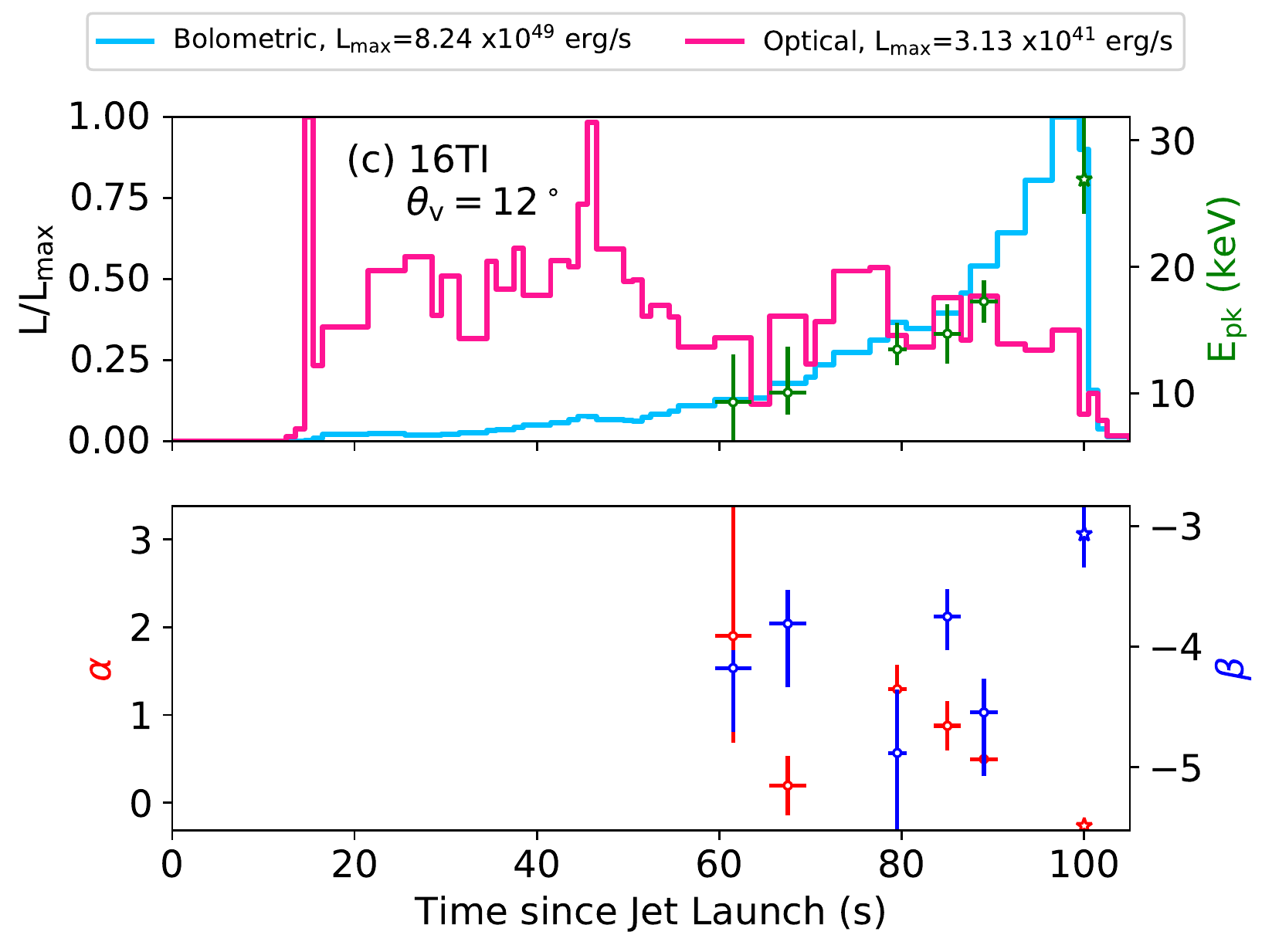}{0.5\textwidth}{\label{16ti_12_lc}}
  }
 \caption{Bolometric and optical light curves and fitted time resolved spectral parameters of the \steady simulation for three observer viewing angles of $\theta_\mathrm{v}=1^\circ, 7^\circ$ and $12^\circ$ in Figures (a), (b) and (c) respectively. The top panels show the bolometric and optical light curves in light blue and pink respectively, each normalized by its maximum luminosity. The green points in the top panel are the fitted time resolved spectral peak energies while the bottom panels show the fitted $\alpha$ and $\beta$ parameters in red and blue respectively. The unfilled markers represent spectra that are best fit with a Band function and solid markers show spectra that are best fit with a COMP spectrum. Star markers represent spectra that are best fit with $\alpha<0$.} 
 \label{16TI_light_curves}
\end{figure*}

\begin{figure}[]
 \centering
 \gridline{
 \fig{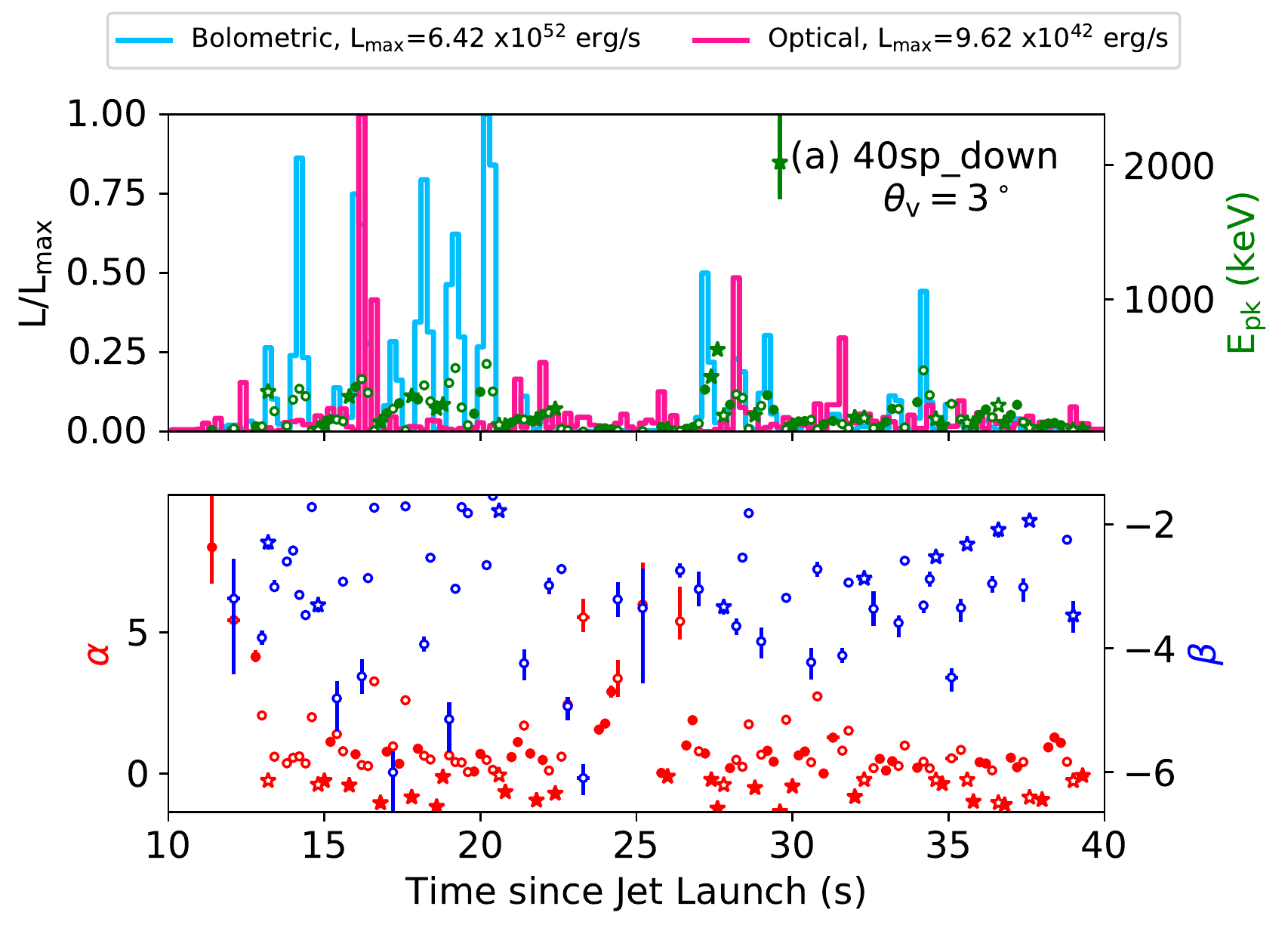}{0.5\textwidth}{\label{40sp_down_3_lc}}
 }
 \gridline{
 \fig{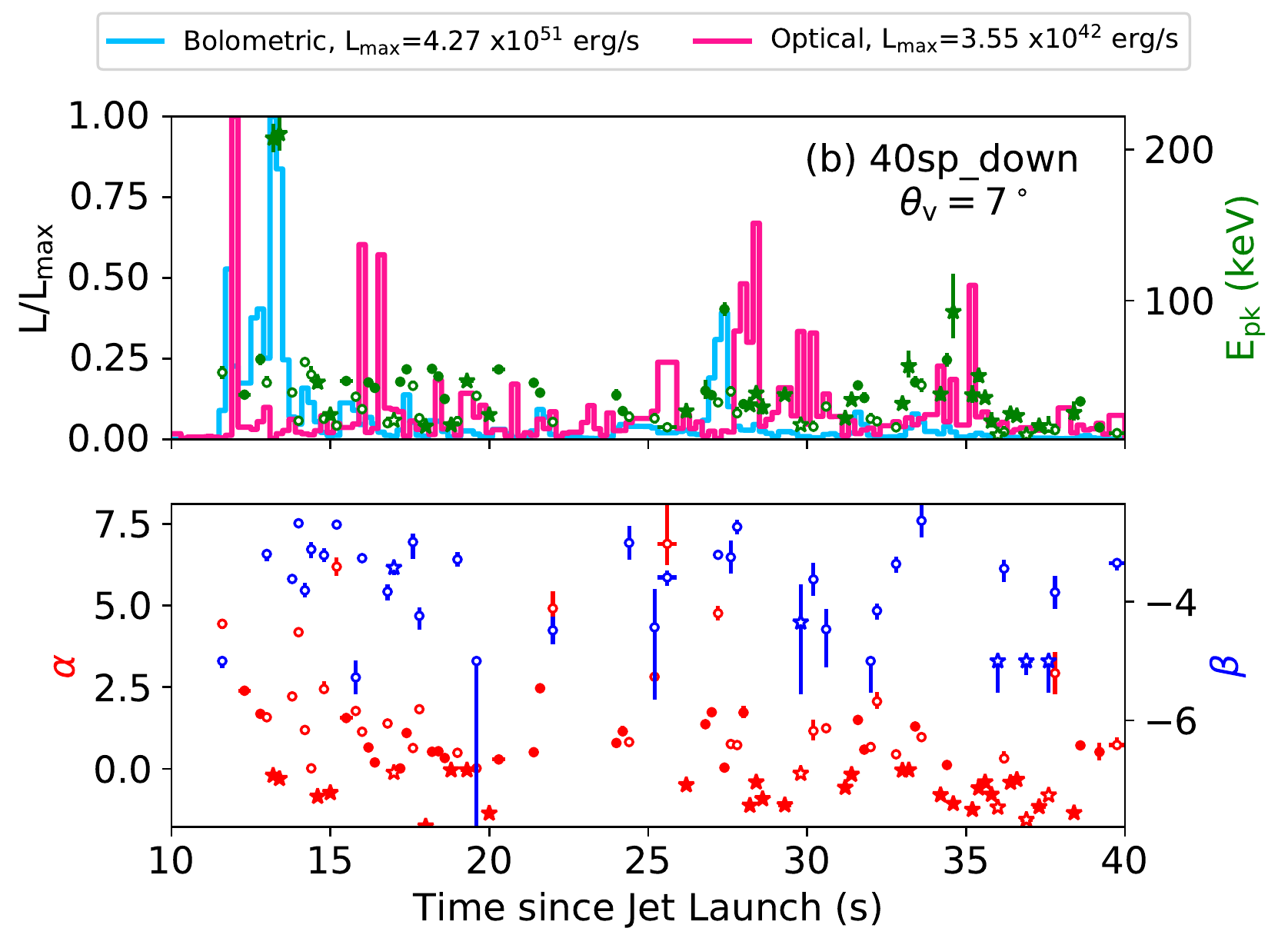}{0.5\textwidth}{\label{40sp_down_7_lc}}
 }
 \caption{The same as Figure \ref{16TI_light_curves} except the mock observations are for the \spikes simulations for $\theta_\mathrm{v}=3^\circ$ and $7^\circ$.} 
 \label{40sp_down_light_curves}
\end{figure}

\begin{figure}[]
 \centering
 \epsscale{1.20}
 \plotone{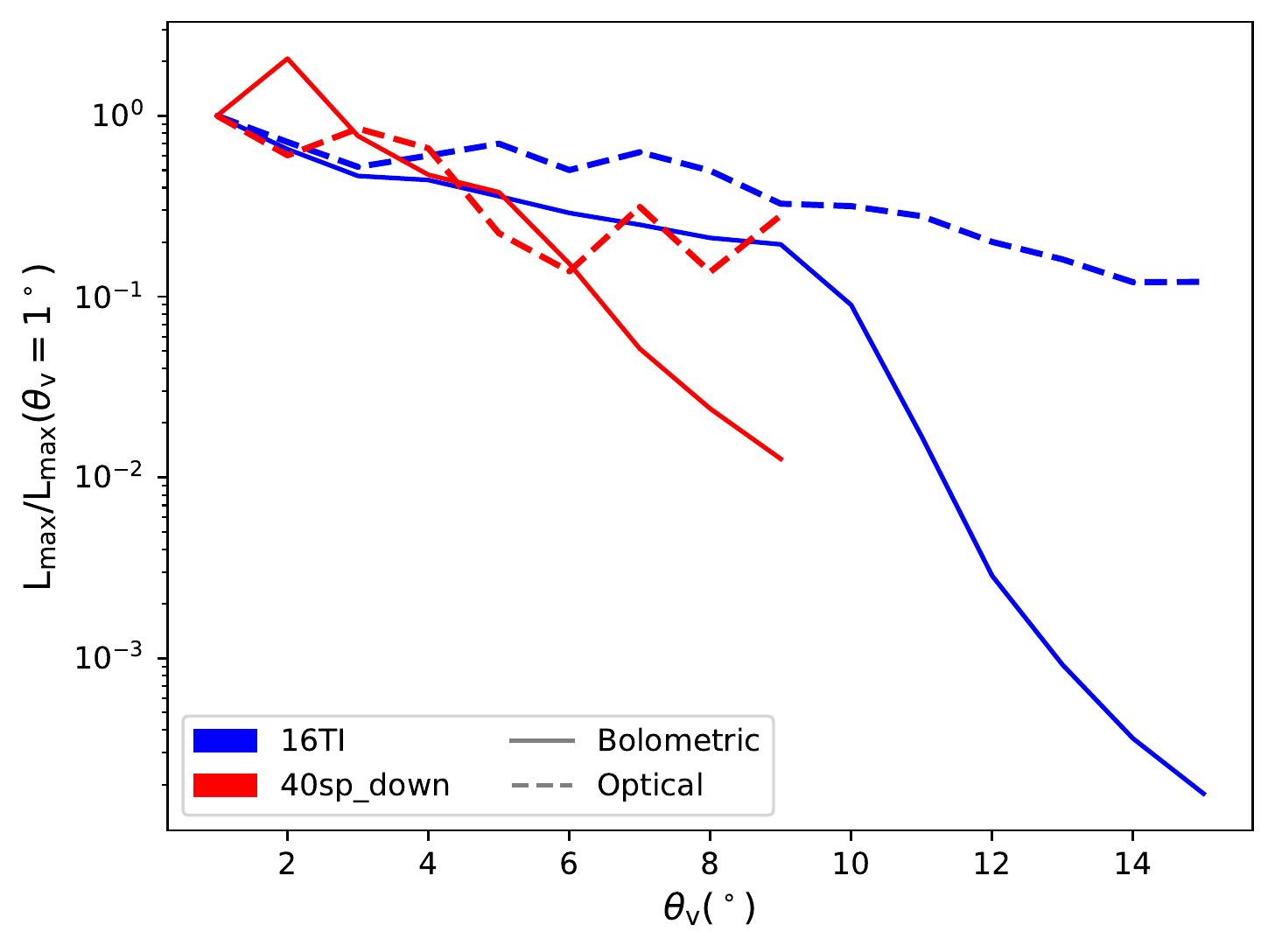}
 \caption{The light curve maximum value, $L_\mathrm{max}$, as a function of observer viewing angle, $\theta_\mathrm{v}$ for the \steady and \spikes simulations, as shown in blue and red respectively. The solid lines represent the  $L_\mathrm{max}$ of the bolometric light curves of each simulation while the dashed lines show the optical $L_\mathrm{max}$ for each simulation.} 
 \label{lumi_bolo_opt_decline}
\end{figure}

Figures \ref{16TI_light_curves} and \ref{40sp_down_light_curves} show mock observed light curves, with variable time bins, of the \steady and \spikes MCRaT simulations, respectively, at various $\theta_\mathrm{v}$. The top panels show the bolometric light curve in light blue normalized by its peak, $L_\mathrm{max}$, the optical light curve in pink, also normalized by its own $L_\mathrm{max}$, and the fitted time resolved peak energies in green. In the bottom panels, the time resolved fitted $\alpha$ and $\beta$ parameters are shown in red and blue respectively; unfilled $\alpha$, $\beta$, and $E_\mathrm{pk}$ markers represent spectra that are best fit by a Band spectrum while solid markers represent spectra best fit with the COMP spectrum. We find that as $\theta_\mathrm{v}$ increases for the \steady simulation, the bolometric light curve, which is dominated by gamma rays, begins to peak at later and later times. For small $\theta_\mathrm{v}$, we find that there can be optical flares after the bolometric light curve has decayed significantly while for large $\theta_\mathrm{v}$ we find that the optical light curves show activity well before the bolometric light curve peaks. In the \spikes case both the optical and bolometric light curves show erratic behavior due to the variability of the injected jet. In both the \steady and \spikes simulations we find that the optical $L_\mathrm{max}$ decays slowly as a function of $\theta_\mathrm{v}$, while the bolometric $L_\mathrm{max}$ decays rapidly, as is shown in Figure \ref{lumi_bolo_opt_decline}. Here, we show the normalized $L_\mathrm{max}$ for the \steady and \spikes optical and bolometric light curves where the normalization is the value of each $L_\mathrm{max}$ at  $\theta_\mathrm{v}=1^\circ$. The blue and red colors represent the \steady and \spikes $L_\mathrm{max}$ while the solid and dashed lines represent the bolometric and optical $L_\mathrm{max}$. 

\begin{figure}[]
 \centering
 \gridline{
 \fig{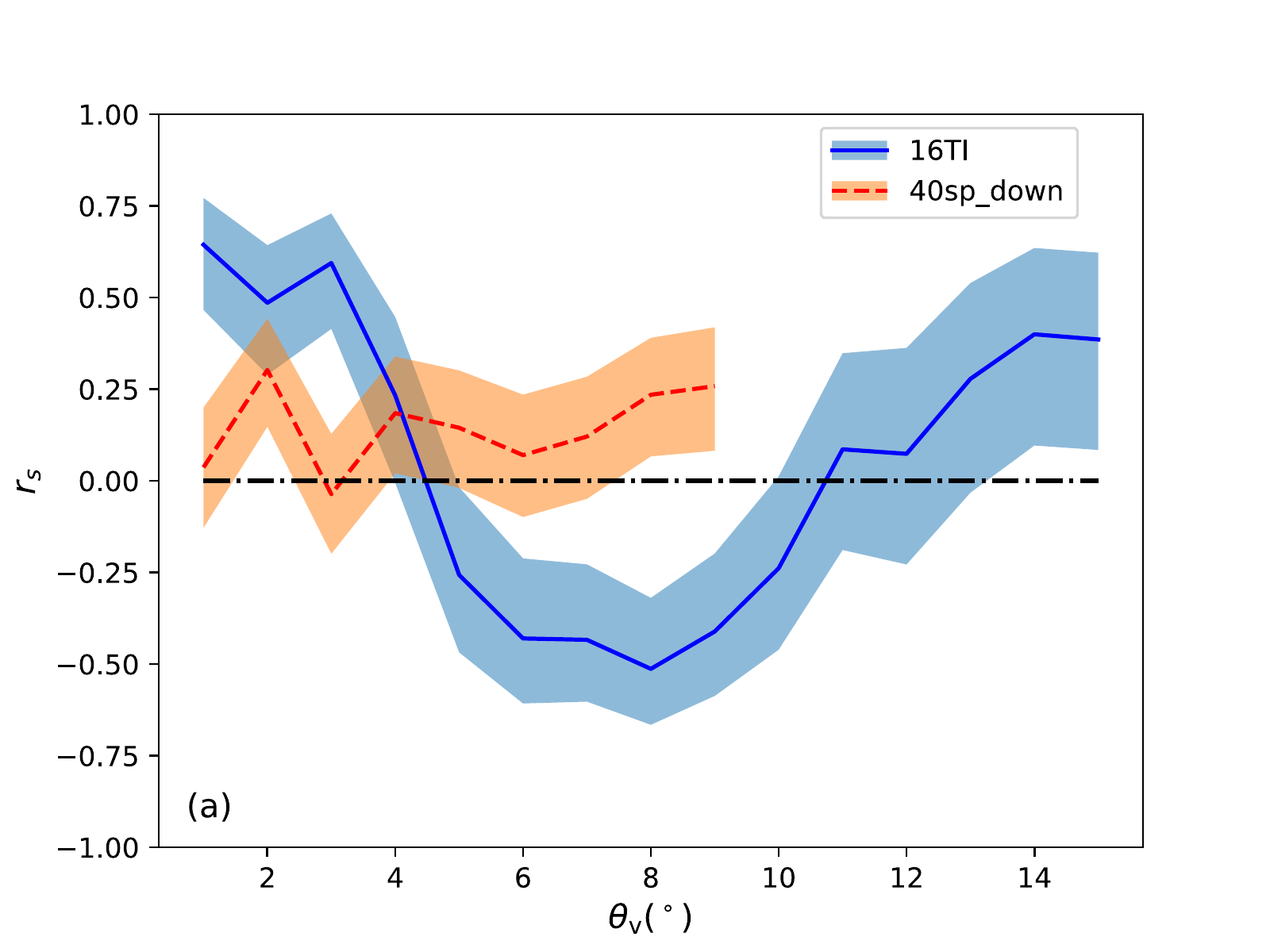}{0.5\textwidth}{\label{spearman}}
 }
 \gridline{
 \fig{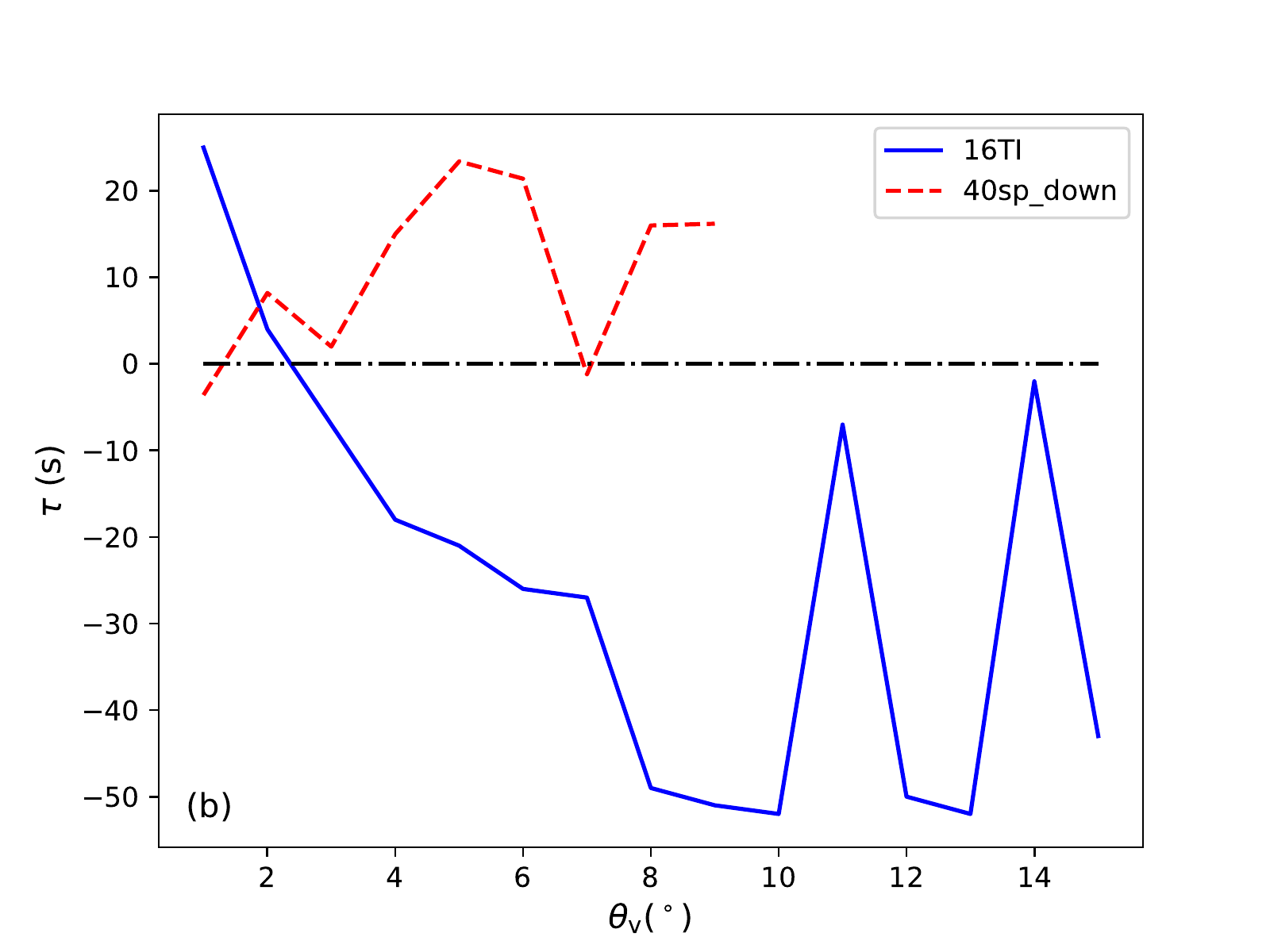}{0.5\textwidth}{\label{time_lag}}
 }
 \caption{Figures (a) and (b) show the Spearman's rank coefficient, $r_s$, and the time lag, $\tau$, between the bolometric and optical light curves of the \steady and \spikes simulations, respectively. We show the results of the \steady simulation with a solid blue line and the \spikes results with a red dashed line. In Figure (a) we also plot the 95\% confidence interval of the calculated $r_s$ with the shaded red and blue regions around each line.} 
 \label{lc_analysis}
\end{figure}

We can analyze the relationship between the bolometric and optical light curves by looking at how well correlated they are through the spearman rank coefficient, $r_s$, and by calculating the time lag, $\tau$, between the two time signals that maximizes their correlation, which we show in Figures \ref{lc_analysis}(a) and \ref{lc_analysis}(b). The red dashed line in each plot shows the results of the \spikes simulation, the blue solid line shows the results of the \steady simulation, and the black dashed-dotted line shows no correlation or time lag. In Figures \ref{lc_analysis}(a), we show the 95\% confidence interval for the calculated $r_s$ through the shaded red and blue regions for the \spikes and \steady simulations respectively. We find that $r_s$ for the \spikes simulation is consistent with 0 for nearly all $\theta_\mathrm{v}$. For the \steady simulation, we find a positive correlation between the bolometric and optical light curves until $\theta_\mathrm{v} \sim 5^\circ$. For $\theta_\mathrm{v} \sim 5^\circ-10^\circ$ we see that the two light curves are negatively correlated and for  $\theta_\mathrm{v} \gtrsim 10^\circ$ the two light curves are marginally correlated. Looking at $\tau$, we find that the \steady simulation shows an evolution from positive time lags, meaning that the optical $L_\mathrm{max}$ occurs later in time than the $L_\mathrm{max}$ of the bolometric light curve, to negative time lags, meaning that the optical $L_\mathrm{max}$ occurs earlier in time than the bolometric $L_\mathrm{max}$ (see also the light curves shown in Figure \ref{16TI_light_curves}). The \spikes $\tau$ are typically positive, showing that the optical $L_\mathrm{max}$ occurs later in time than the bolometric $L_\mathrm{max}$.

\begin{figure*}[]
 \centering
 \gridline{
 \fig{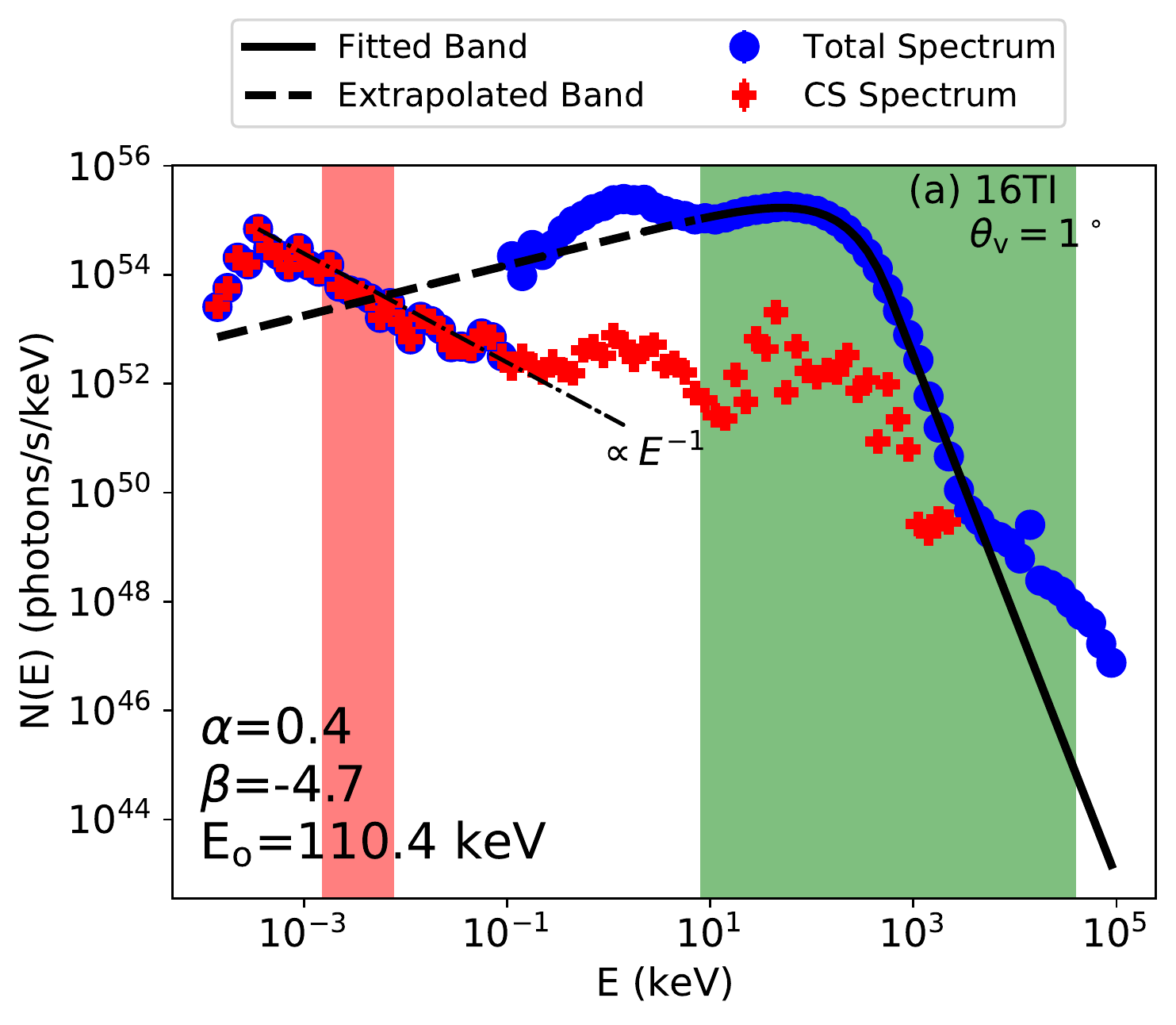}{0.5\textwidth}{\label{16ti_1_spex}}
 \fig{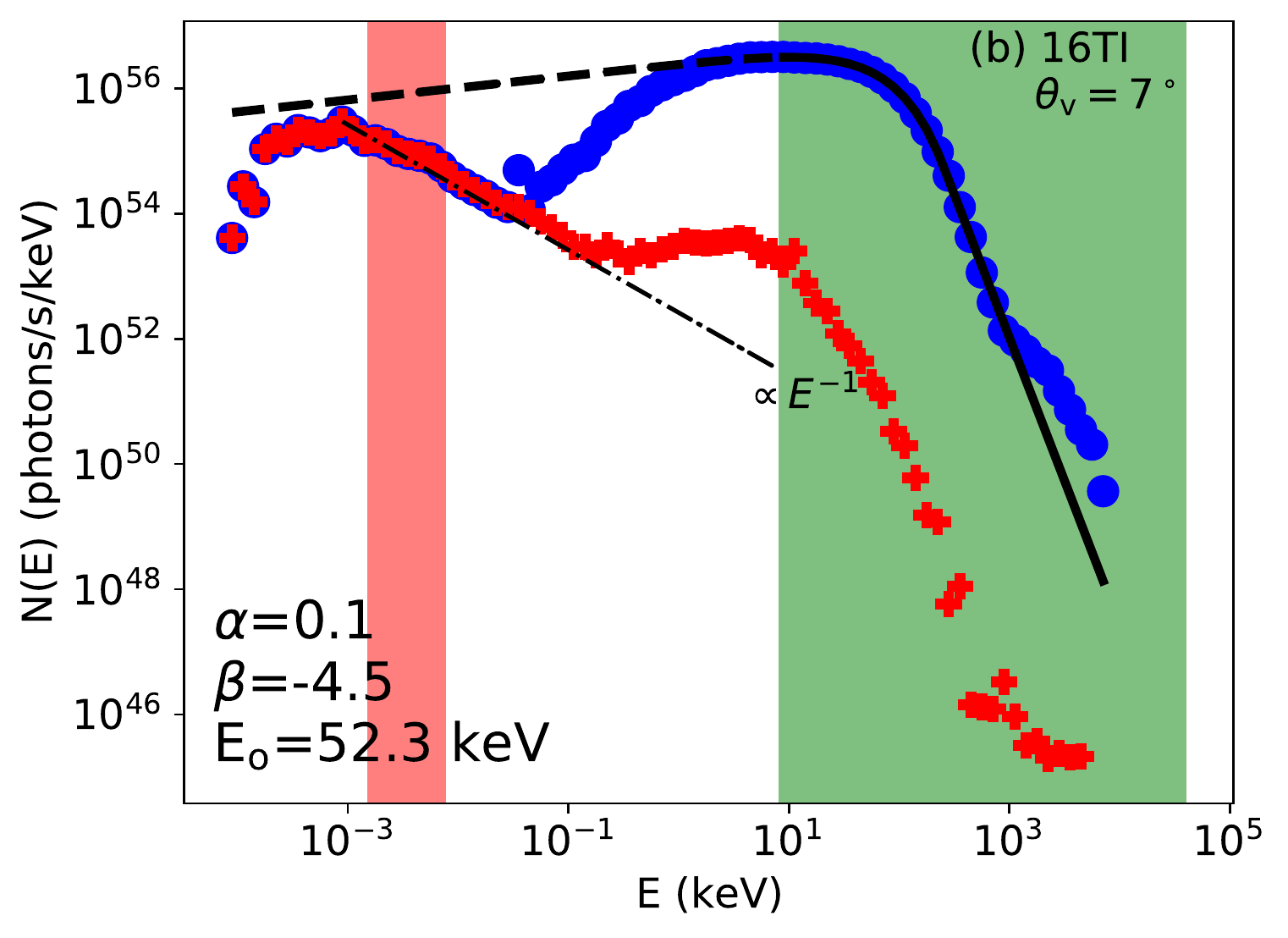}{0.5\textwidth}{\label{16ti_7_spex}}
 }
 \gridline{
  \fig{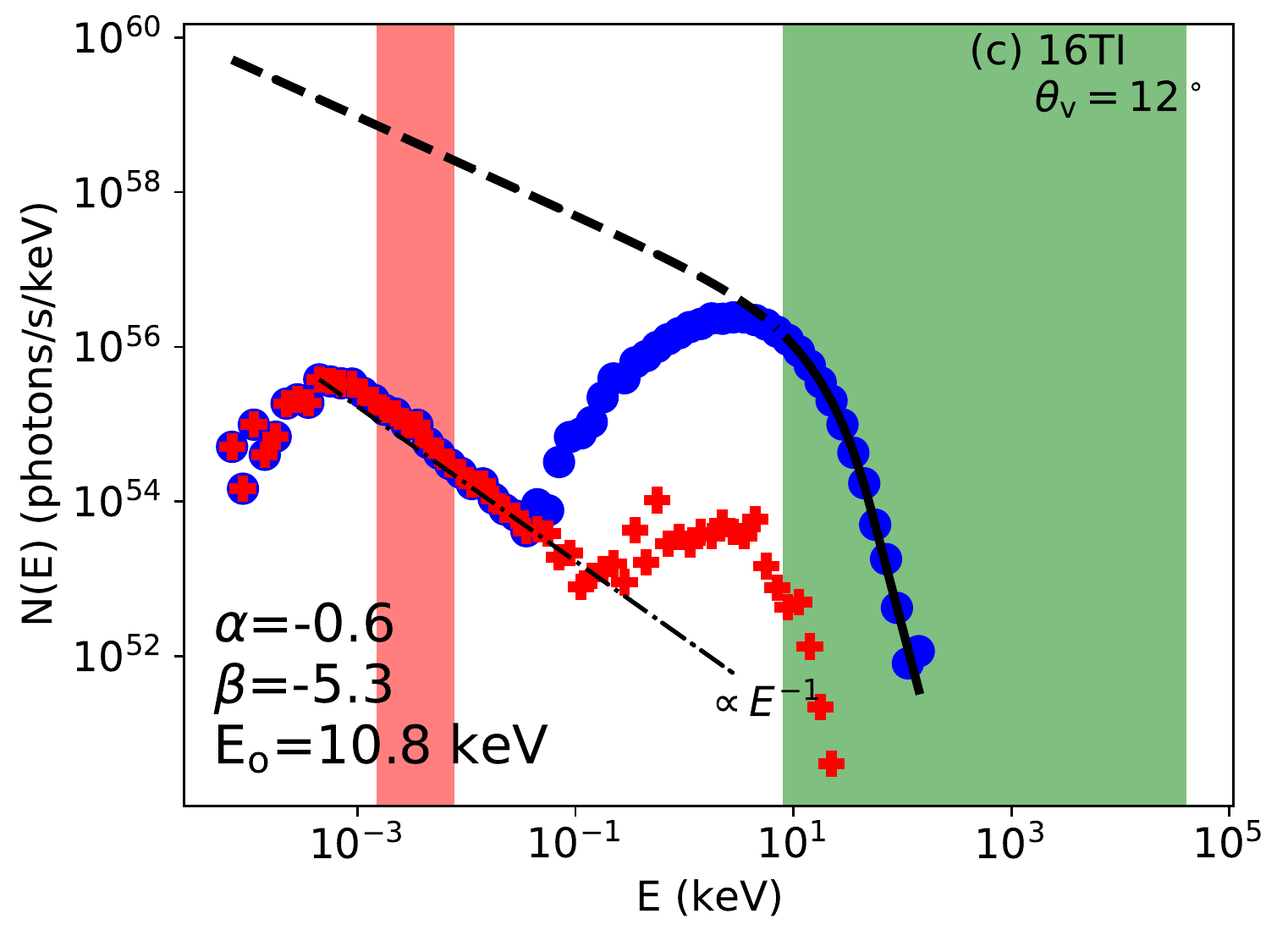}{0.5\textwidth}{\label{16ti_12_spex}}
  }
 \caption{Time integrated spectra for the \steady simulation for the same observer viewing angles presented in Figure \ref{16TI_light_curves}. The blue circle markers represent the total spectrum consisting of the thermally injected photons and the CS photons. The red plus markers show the spectrum of just the CS photons. The pink shaded area is the bandpass of the Swift white filter and the green shaded area corresponds to the bandpass of the Fermi telescope. The fit to the Fermi portion of the spectra is shown by the solid black line, with the fitted parameters given in the bottom left corner, while the extrapolated portion of the fitted spectrum to optical wavelengths is shown by the dashed black line. The thin dashed-dotted black line shows the slope of $N(E)\propto E^{-1}$. {{In Figure (c) we also plot in black the observed optical prompt emission spectral photon number density for GRBs 080319B and 041219A in relation to the MCRaT spectra.}} } 
 \label{16TI_spectra}
\end{figure*}

\begin{figure}[]
 \centering
 \gridline{
 \fig{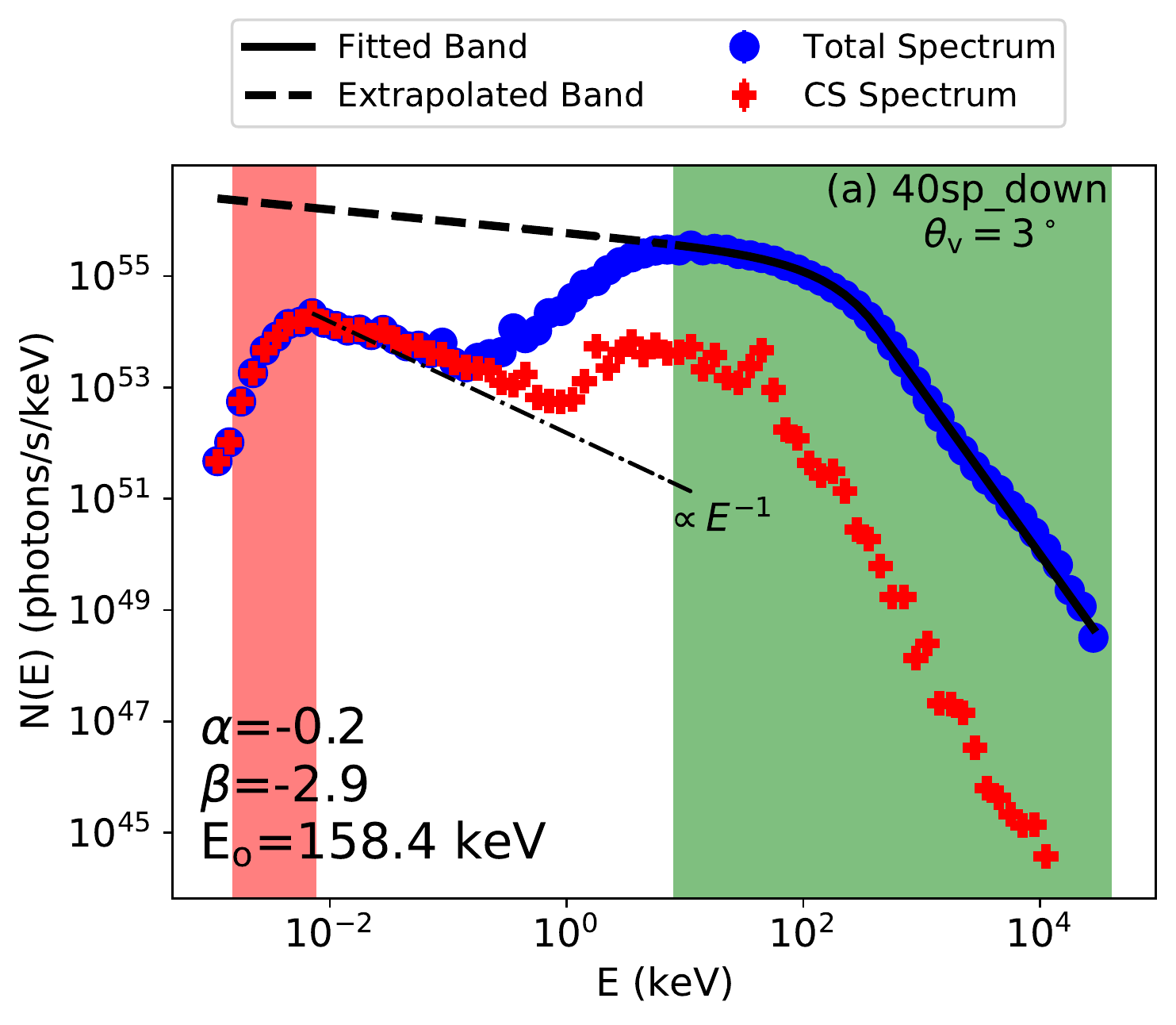}{0.5\textwidth}{\label{40sp_down_3_spex}}
 }
 \gridline{
 \fig{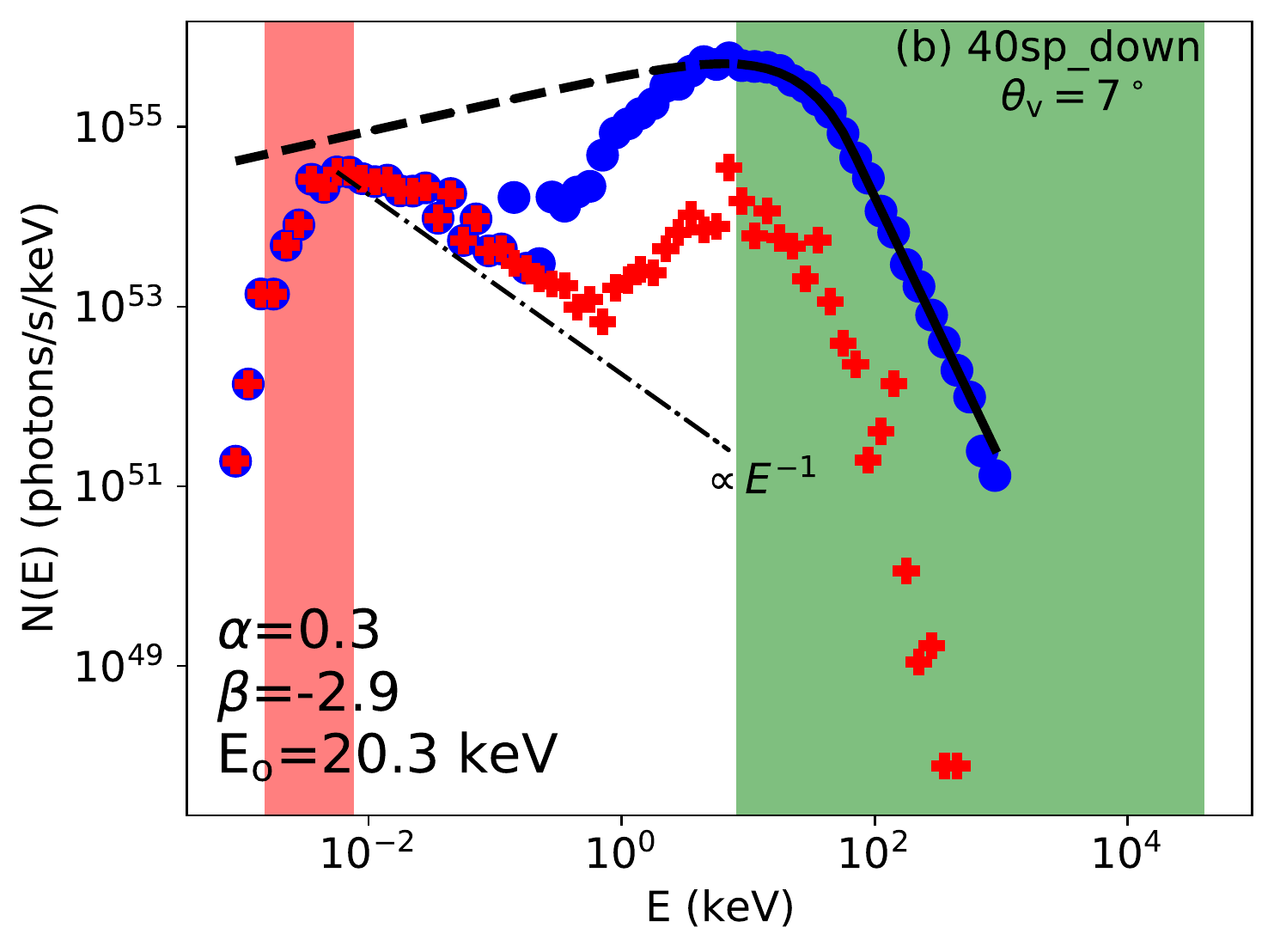}{0.5\textwidth}{\label{40sp_down_7_spex}}
 }
 \caption{Same as Figure \ref{16TI_spectra} except for the \spikes simulation at $\theta_\mathrm{v}=3^\circ$ and $7^\circ$.} 
 \label{40sp_down_spectra}
\end{figure}

\begin{figure}[]
 \centering
 \gridline{
 \fig{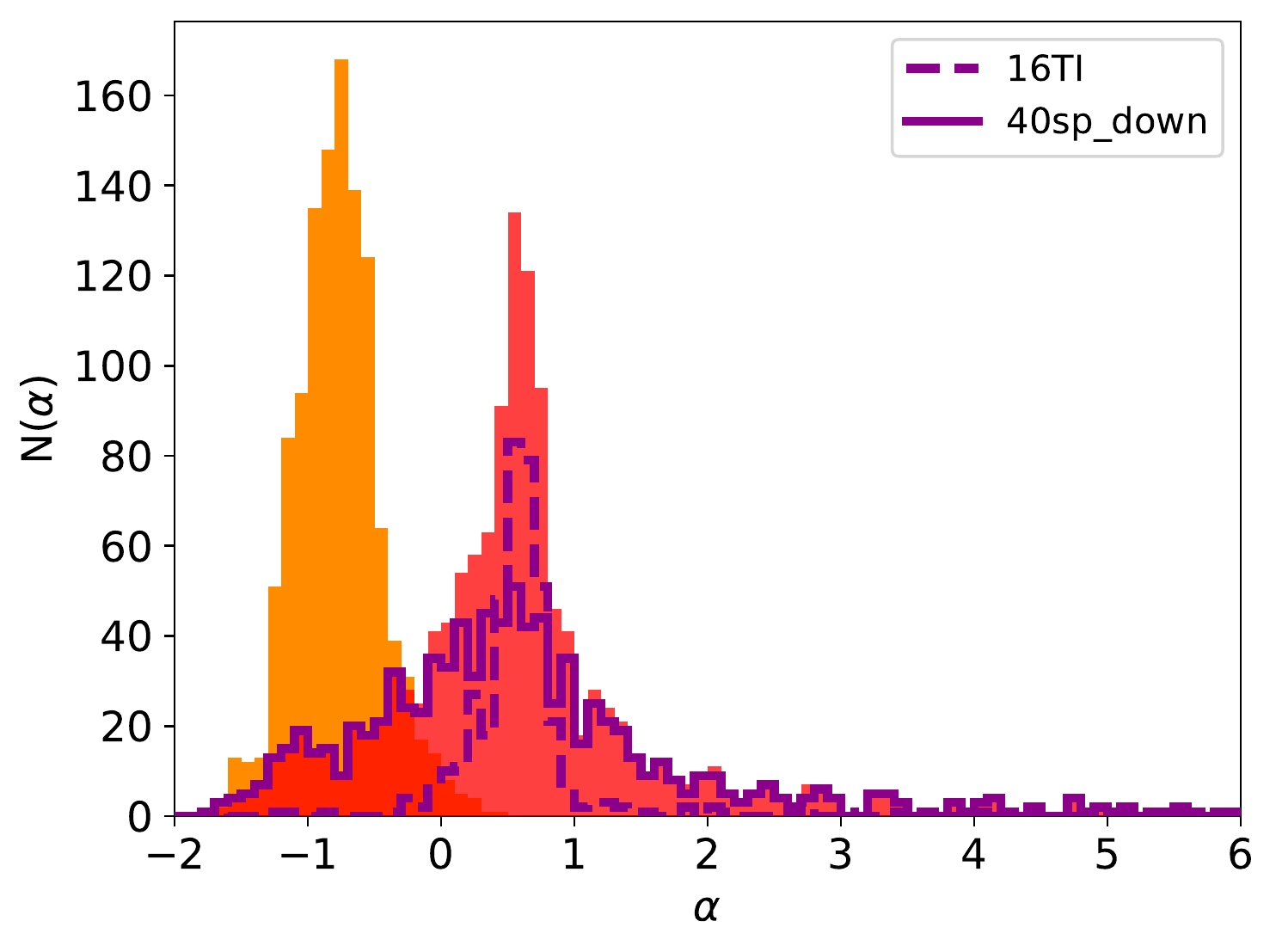}{0.45\textwidth}{\label{alpha}}
 }
 \gridline{
 \fig{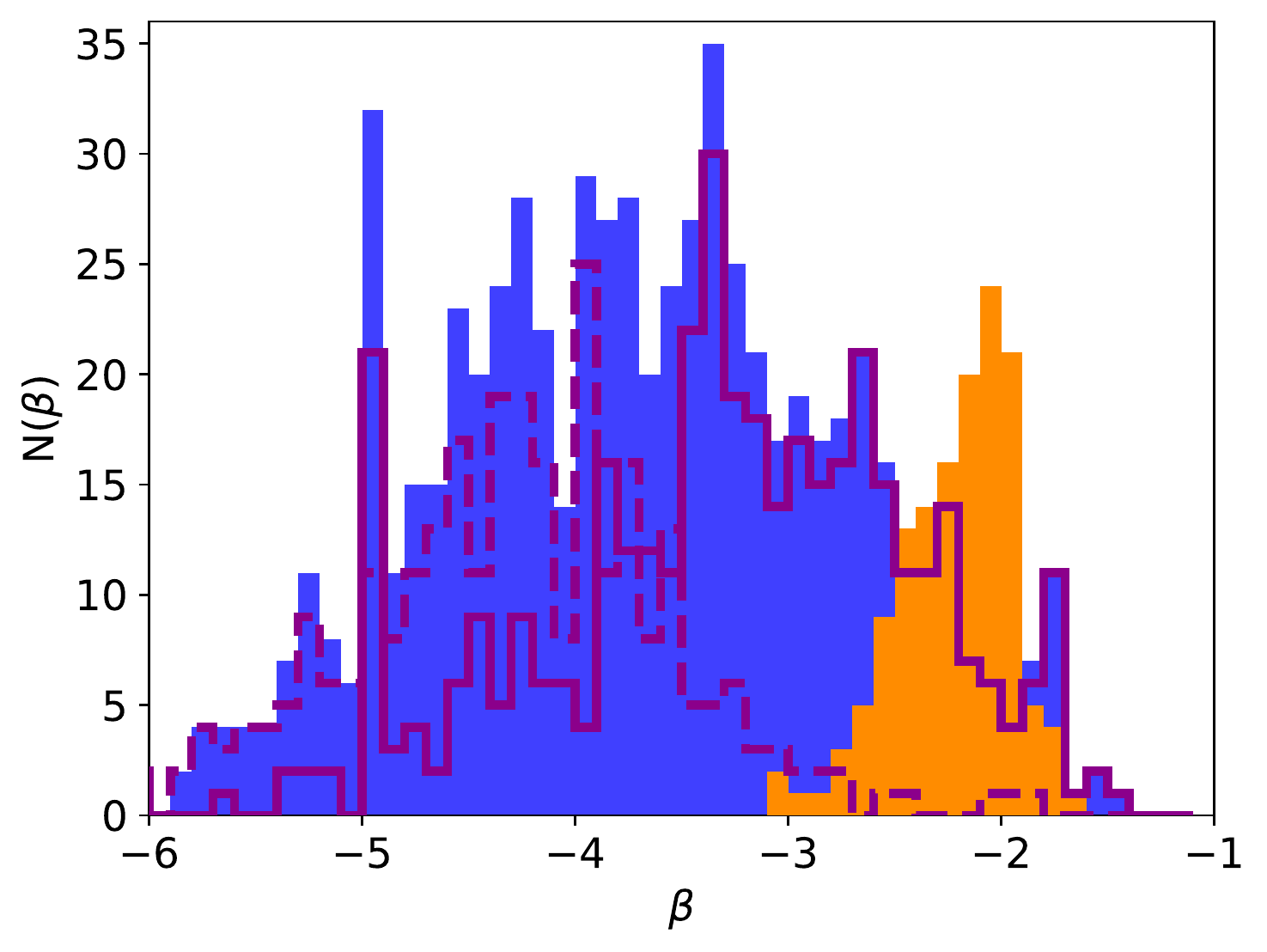}{0.45\textwidth}{\label{beta}}
 }
 \gridline{
  \fig{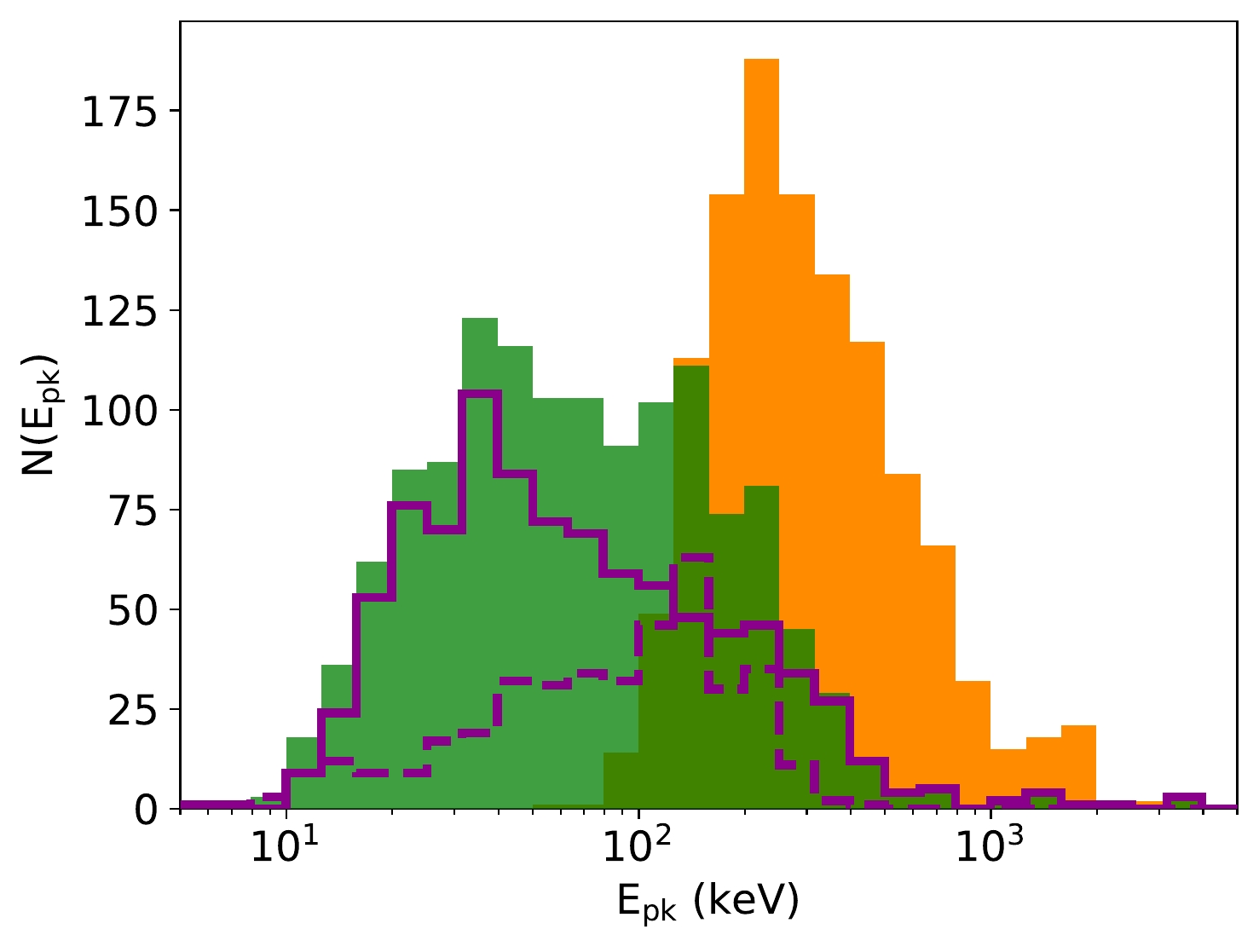}{0.45\textwidth}{\label{e_pk}}
  }
 \caption{Histograms of the fitted time-resolved MCRaT spectral parameters for the \steady and \spikes simulations in red, blue, and green. The orange histograms are the fitted observed spectral parameters from \cite{FERMI}. The purple lines denote the contributions from the \steady simulation, with the dashed line, or the \spikes simulation, with the solid line.} 
 \label{spectral_fits}
\end{figure}

\subsection{Time Resolved and Time Integrated Spectra} \label{section_spectra}
In this section, we discuss the results of our mock observed MCRaT spectra. Figures \ref{16TI_spectra} and \ref{40sp_down_spectra} show time integrated spectra for the \steady and \spikes simulations, respectively, for the same $\theta_\mathrm{v}$ shown in Figures \ref{16TI_light_curves} and \ref{40sp_down_light_curves}. The total spectra, consisting of all the detected photons, are shown as blue circle markers while the CS spectra, consisting of just the scattered CS photons, are represented with red plus markers. The total spectra are fitted with either a Band or COMP spectrum within the green shaded region, which represents the photon energies that are typically considered in observational analysis of Fermi GRB spectra. The solid black line denotes the best fit function with its best fit parameters given in the bottom left of the plots and the dashed black line shows the extrapolation of the fit to lower energies. The red shaded region highlights the Swift white filter bandpass, which we use to produce our optical light curves in the previous section. {{We have also plotted the spectral photon density of the optical prompt emission of GRBs 080319B and 041219A in relation to the MCRaT spectra in Figure \ref{16TI_spectra}(c).}}

In the \steady simulation, we find that the time integrated spectra are best fit with positive $\alpha$ for $\theta_\mathrm{v}\le 8^\circ$, while for $\theta_\mathrm{v}>8^\circ$, the portion of the spectra in the Fermi band is best fit with a negative $\alpha$, reproducing typical observed GRB time integrated spectral $\alpha$. For the \spikes simulation, we observe the opposite trend, where for $\theta_\mathrm{v}\le 5^\circ$ the spectra are best fit with a negative $\alpha$ while at larger $\theta_\mathrm{v}$ the fits utilize a positive $\alpha$ and diverge from real observations of GRBs. In both cases we find that below $\sim 0.1$ keV the spectra are composed of the CS photons primarily; above this energy, the CS photons do not affect the spectra significantly. The CS photons in both simulations also form a powerlaw of $\sim  E^{-1}$ between $\sim 10^{-3}$ keV or $\sim 10^{-2}$ keV to $\sim 0.1$ keV. Looking at the extrapolated fitted spectra and the intensity of the spectra at optical energies, we find a variety of behaviors from the optical emission lying above or below the extrapolated fits. In particular, time integrated spectra with $\alpha <0$ have optical emission that lie far below that of the extrapolated Band function at these energies. {{Another important result is that the optical region of the MCRaT spectra lie far below what is observed, regardless of the simlation type or spectral fit.}}

We can also look at the MCRaT time resolved spectral fits and compare them to typical observational GRB spectral fits and their distributions, as is shown in Figure \ref{spectral_fits}. In Figure \ref{spectral_fits} we plot the distributions of $\alpha$, $\beta$, and $E_\mathrm{pk}$, shown in red, blue, and green respectively, alongside distributions from \cite{FERMI}. The purple solid and dashed lines represent the contributions to the histograms from the \spikes and \steady simulations respectively. We find that the fitted time resolved $\alpha$ parameters are widely distributed around $\sim 0.5$ with a skewed distribution to negative values which is made up of spectra from the \spikes simulation. The $\beta$ values are also widely distributed with values that are steeper than what is found by \cite{FERMI}. The MCRaT $E_\mathrm{pk}$ distribution is skewed towards lower values than those found by \cite{FERMI}, mostly due to the low luminosity points of the \spikes simulation, where photon energies are much lower than expected \citep{parsotan_var}. These results differ slightly from prior analysis due to the fact that we fit the spectra from 8 keV to 40 MeV and, as a result, more spectra are best fit with negative $\alpha$ values, showing the importance of taking observational effects into account.

\subsection{Location of Photons in the Jet}
\begin{figure*}[]
 \centering
 \fig{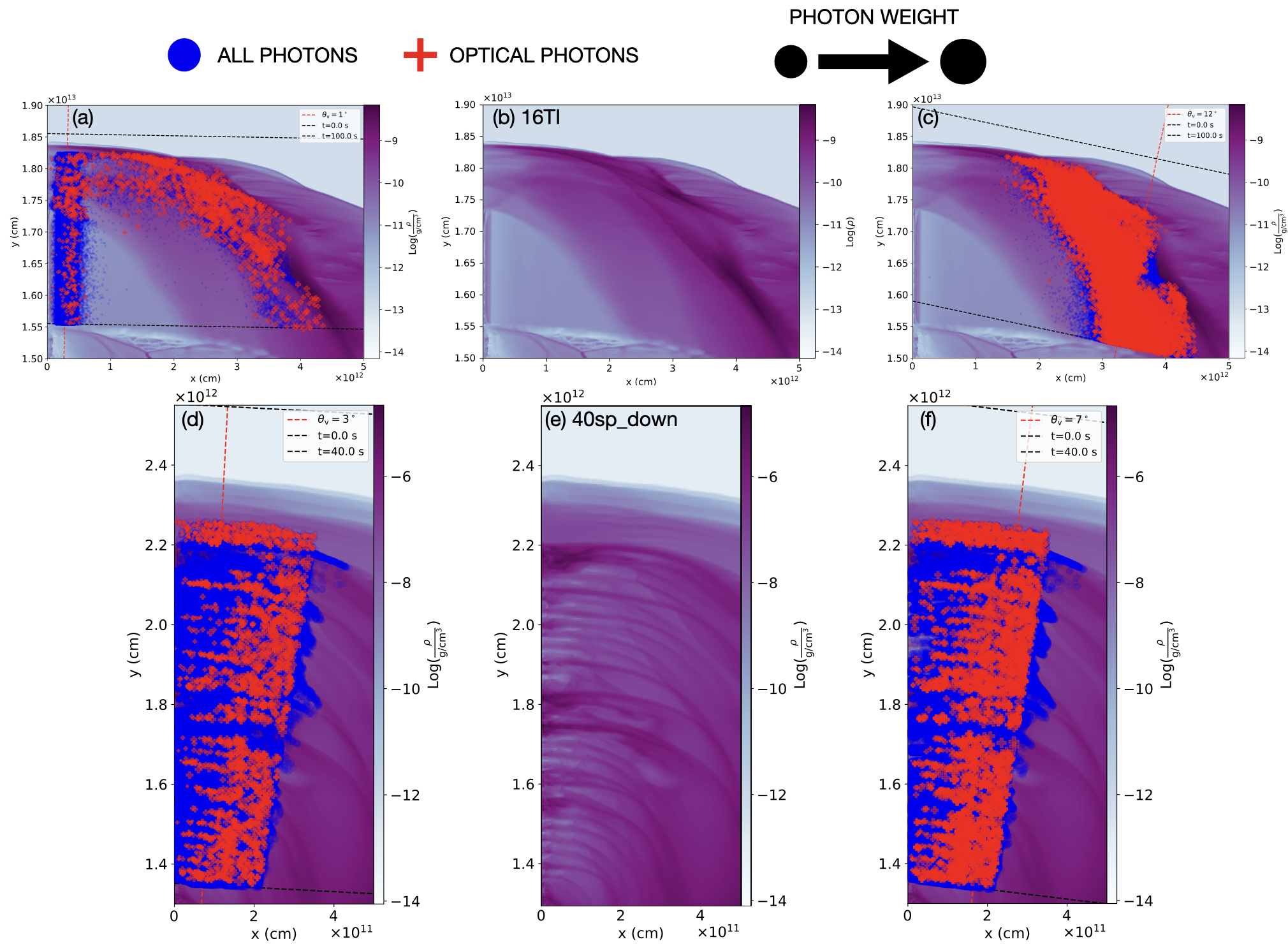}{\textwidth}{}
 \caption{Pseudocolor density plots of the \steady and \spikes simulations with the location of all the photons that an observer, located at $\theta_\mathrm{v}$, would detect over the entirety of the light curve shown in either Figure \ref{16TI_light_curves} or \ref{40sp_down_light_curves}. The red dashed line shows the direct line of sight of the observer located at $\theta_\mathrm{v}=1^\circ, 12^\circ, 3^\circ$, and $7^\circ$ in the \steady and \spikes simulations, as is shown in Figure (a), (c), (d), and (f) respectively. The equal arrival time surfaces (EATS) for the start and end of the light curves are shown as black dashed lines. Each photon that is detected is shown with semi-transparent blue circles and the optical photons are shown in semi-transparent red plus markers. As more photons are located in a given region of the outflow, the color density of each type of photon increases. The size of each marker denotes the photon weight, with photons represented by smaller markers having less weight, and thus making a smaller contribution to the calculation of light curves and spectra, and vice versa for photons represented with large markers. Figures (b) and (e) show only the pseudocolor density plot in order to identify features in the outflows that can be related to where the photons are located. } 
 \label{photon_location_fig}
\end{figure*}
In order to get a deeper understanding of what the mock observations presented in Section \ref{section_lc} and \ref{section_spectra} tell us about what is occurring in the simulated GRB jet, we plot the location of the detected photons in relation to the SRHD jet.

Figure \ref{photon_location_fig} shows the pseudocolor density plots of the \steady and \spikes simulations, in the top and bottom rows respectively. The middle figures ((b) and (e)) show just the pseudocolor density plots while the remaining plots show the location of the photons that are detected by an observer at the given $\theta_\mathrm{v}$. These photons are used to construct the light curves shown in Figures \ref{16TI_light_curves} and \ref{40sp_down_light_curves}. The observers line of sight direct to the location of the central engine is shown by the red dashed line and the black dashed lines show the EATS corresponding to $t=0$ s in the light curves and the end time of the light curves. The semi-transparent blue circle markers show all the photons that the observer detects regardless of energy while the semi-transparent red plus markers show the photons with energies in the Swift white band. The size of the markers also denote the weight of each photon; since the weight represents the number of physical photons each MCRaT photon packet represents, this exhibits the contribution that a given photon makes in the calculation of the mock observables presented in Section \ref{section_lc} and \ref{section_spectra}. If many photons are located in a given area of the outflow the color density of blue or red photons increases, while photons with larger weights in the outflow are shown by larger markers. 

Looking at Figure \ref{photon_location_fig}(a) where $\theta_\mathrm{v}=1^\circ$, we see that for an observer located along the jet axis all the photons that would be observed are either located along the jet axis or along the shocked interface of the jet and the cocoon (the jet cocoon interface or JCI; \cite{gottlieb2021structure}). Furthermore, when looking at the optical photons, we find that the photons along the JCI have larger weights especially at late times in the light curve (see Figure \ref{16TI_light_curves}(a)), showing that more optical photons are coming from this region of the jet than from the stream of fluid directly along the observer's line of sight. For an observer located further from the jet axis at $\theta_\mathrm{v}=12^\circ$, as is shown in Figure \ref{photon_location_fig}(c), the emission that is detected is entirely from the JCI. At early times, the emission is from the JCI that is near the core of the jet, which is when the bolometric light curve is at a minimum and the optical light curve is at a maximum (see Figure \ref{16TI_light_curves}(c)). At late times, the detected photons are located along the observers line of sight, corresponding to a minimum in the optical light curve and a maximum in the bolometric light curve.

Figures \ref{photon_location_fig}(d) and (f) show the locations of the observed photons in the \spikes simulation for observers located at $\theta_\mathrm{v}=3^\circ$ and $7^\circ$, respectively. In this simulation, we find that photons, regardless of their energy, are being detected by the observer from nearly all regions of the jet. This is expected since the jet is moving with a low Lorentz factor \citep{parsotan_var}, allowing photons from high latitude regions of the outflow to scatter into the observers line of sight. If we focus on the optical photons, we find that they originate mostly from high latitude regions of the jet, corresponding to the JCI, and where shocks occur in the jet. The emission regions of the optical emission nearly trace out the shocks in the \spikes simulation, showing the importance of taking these shocks and the JCI into account in radiative transfer calculations.

\section{Summary and Discussion} \label{summary}
We have used the MCRaT simulations in order to analyze two SRHD simulations of GRB jets which we denote the \steady and \spikes simulations. The MCRaT code has been improved to include cyclo-synchrotron (CS) emission and absorption which has been tested and verified with other works in the literature. As a result of this improvement, we are now able to make predictions of expected light curves and spectra under the photospheric model of GRBs from optical to gamma ray energies. 

Our results can be summarized as follows:
\begin{itemize}
\item Jets with a constant injection of energy, such as the \steady simulation, show a Spearman rank correlation between their optical and bolometric light curves (which is dominated by gamma rays) that changes from being positively correlated until $\theta_\mathrm{v} \sim \theta_o/2$, where $\theta_o$ is the jet opening angle. From $\theta_\mathrm{v} \sim \theta_o/2$ to $\sim \theta_o$, there is a negative correlation between the two time series and after  $\theta_\mathrm{v} \sim \theta_o$ there is a marginal positive correlation. In the \spikes simulation, indicative of variable GRB jets, we find that there is no significant correlation at almost all $\theta_\mathrm{v}$ and that the optical emission typically lags the gamma ray emission event.

\item The \steady simulation's time lag shows that as $\theta_\mathrm{v}$ increases, there is expected to be optical precursors up to a few tens of seconds before the main gamma ray emission.

\item The time integrated fitted Band spectra, in the Fermi energy band from 8 keV - 40 MeV, of our simulations can be best fit with negative $\alpha$ parameters. We also get an increased number of time resolved spectra, particularly in the \spikes simulation, that are best fit with a negative $\alpha$.

\item The CS photons do not have a large effect on the peak of the spectrum, however CS emission produces optical photons that exhibit a power law of $\sim E^{-1}$ below $\sim 0.1$ keV. 

\item In time integrated spectra that are best fit with negative $\alpha$ parameters, the optical emission lies below the extrapolated Band function.

\item For the \steady simulation, we found that an observer located along the jet axis sees photons directly along their line of sight in the jet in addition to emission from the Jet Cocoon Interface (JCI). The detected optical photons, for an observer located along the jet axis, mostly originate from the JCI. For observers located far from the jet axis, the detected photons (optical to gamma-rays) originate solely from the JCI. In this case, the detected photons follow the surface of the JCI, with photons detected early on originating close to the jet core and photons detected later originating from the JCI that is along the observer's line of sight.

\item In the \spikes simulation, photons are detected from all regions of the jet due to the low Lorentz factors present in the outflow. The optical photons are located at the interfaces of shocks in the outflow and at high latitude regions of the JCI.
\end{itemize}

In line with our result that CS emission does not change the spectral peak energy of the synthetic spectra, we also found that the location of the \steady and \spikes simulations in the Amati, Yonetoku, and Golenetskii relationships were consistent with the prior results of \cite{diego_lazzati_variable_grb}, \cite{parsotan_mcrat}, and \cite{parsotan_var}, where the simulations are in agreement with the Yonetoku and Golenetskii relationships and in tension with the Amati relationship.

Our findings have observational consequences for observing the prompt emission. The MCRaT simulations have shown that there can be various correlations between the optical and gamma ray energy light curves depending on $\theta_\mathrm{v}$ and the intrinsic variability of the GRB jet. Traditionally, optical emission has been attributed to the emission mechanism that produces the prompt emission in gamma rays due to temporal correlations between the two signals. We have shown that this is indicative of $\theta_\mathrm{v}$ being close to the GRB jet axis. Furthermore, for GRBs observed far from the jet axis, there is expected to be a negative or no correlation between the optical and gamma ray light curves. The MCRaT results also predict there to be optical prompt precursors occurring for tens of seconds before the main gamma ray emission. Detecting these optical precursors, if they are bright enough, may allow the community to improve followup capabilities and thus be better prepared to measure the prompt and afterglow emission. Another interesting alternative to detecting GRB prompt emission is the possibility of detecting solely the optical prompt emission and not the gamma ray prompt emission. This would be possible at large $\theta_\mathrm{v}$ where the intensity of gamma rays is low enough that current detectors would not trigger. Unlike the gamma ray luminosity declining steeply with $\theta_\mathrm{v}$ \citep{parsotan_polarization}, the optical luminosity is a weak function of $\theta_\mathrm{v}$ which means that it is possible that it would be detected without the gamma ray counterpart, possibly resulting in emission similar to what was observed in PTF11agg \citep{cenko2013discovery}. In these cases, it would be important to consider the effects of the afterglow forward and reverse shocks and understand when the radiation from these processes may begin to wash out the true optical prompt emission, a task that is outside the scope of this paper. Due to our MCRaT optical predictions being due to CS emission and absorption, which is certainly not the only process producing photons with optical energies, the optical luminosities and spectral intensities that we acquire in this paper provide lower limits on the predicted optical prompt emission. We can estimate the detected magnitude of our best case optical prompt emission signature, $L_\mathrm{max}=1.1\times 10^{43}$ erg/s, by assuming a detection with the Swift white filter for a GRB located at redshift $z=1$ with the aforementioned optical prompt luminosity at its maximum. This provides a detected magnitude of $\sim 25$ in the Vega system, which is still a few magnitudes dimmer than some of the dimmest detections made (see Table \ref{opt_reference_table}). {{Furthermore, this estimated magnitude is still many magnitudes dimmer than optical prompt detections made for GRBs located at $z\sim 1$. Thus, detecting optical prompt precursors and measuring correlations between the optical and gamma-ray prompt emissions under the current unmodified photospheric model would be difficult. These predictions may remain the same or change as the photospheric model is modified and additional radiation mechanisms are added to the model to account for typical optical prompt detections, although this will be the subject of future work.}}  

{{Our results are not able to fully account for the optical prompt detections that have been made, showing that, at least in those cases, another emission mechanism needs to be invoked to explain the optical brightness.. Nonetheless, the MCRaT results are able to reinforce some of the results acquired from analysis of optical prompt emission under the synchrotron model. Namely, we find that the optical prompt emission comes from a different region of the jet than the gamma rays typically originate (particularly for $\theta_\mathrm{v}$ close to the jet axis) which is a similar conclusion that \cite{Zou_2009_optical_different_region} reached in thier study. Furthermore, \cite{Tang_2006_optical_lag} find that the optical lag of GRB 041219A is $\sim 1-5$ s which, under our results, suggests that this GRB was observed at $\theta_\mathrm{v}\sim 2^\circ$ regardless of jet model. Finally, analysis of optical prompt emission typically place the emission radius at radii much larger than typical photospheric radii. As we have shown in this work, the photospheric model produces photons in the dense, slow moving JCI for which we can calculate the location of the photospheric radius, $R_{ph}$, at these high latitude regions. These calculations show that $R_{ph}\sim 10^{13}-10^{16}$ cm which is larger than typical photospheric radii for gamma ray emission and aligns with the findings of \cite{shen_synch_optical_emission}.}}

Although the large majority of the mock observed time integrated and time resolved spectra suffer from a lack of low energy photons which contributes to the excess of positive $\alpha$ fits that we acquire, many spectra can be fit with negative $\alpha$ values due to the fact that we impose the observational constraint of fitting the spectra from 8 keV - 40 MeV. This shows the importance of taking observational effects into account when comparing theoretical results to observational results, which should become the standard practice in the community. In order to make mock observations more realistic, we would need to also consider the transmission functions for various optical filters as well as the instrument response functions for various gamma ray instruments.  Including these effects may change the results of the spectral fits, potentially bringing the theoretical spectra into agreement with observed spectra. Additionally, including these effects will allow the community to determine how well observational analyses are able to recover the ground truth of simulations. These instrumental effects and the ability of observational analysis to recover the ground truth of simulations will be the topics of future papers. 

At energies less than 8 keV, the spectra do not reproduce the findings of \cite{Oganesyan2019_prompt_opt} who find that the photon index of the spectra of their sample set is $-2/3$ between the optical and soft X-rays for the synchrotron spectra that they fit to the data and $\sim -1$ for their two component model, which is similar to a Band function at low energies. This discrepancy drives the need for another radiation mechanism in MCRaT. \cite{Oganesyan2019_prompt_opt} calculated the properties of the emitting region based on the superior fits that their synchrotron spectra provided over the two component model and found values for the comoving magnetic field, the number of emitting electrons, the electron Lorentz factor, the emitting radius, and the bulk Lorentz factor that agreed well with marginally fast cooling synchrotron emission \citep{daigne2011marginally_fastcooling_synch}. Although these values are in tension with standard values of the prompt emitting region, the inclusion of synchrotron emission in this regime in the MCRaT code may help produce the correct spectra {{since it has been shown in this work and others (see e.g \cite{Zhang_E_p_evolution}) that photospheric emission on its own is unable to consistently create spectra with photon indexes of $-1$.}} In such a case, we may expect synchrotron emission to form the low energy ($\lesssim 1$ keV) portion of GRB spectra while compton scattering forms the portion of the spectra at high energies, above typical GRB $E_\mathrm{pk}$. In the energy region below $E_\mathrm{pk}$ and above $\sim 1$ keV we would then expect both processes to contribute nearly equally to the spectrum, producing spectra with $\alpha \sim -1$. 

In addition to adding synchrotron physics to MCRaT, we have also shown that we need to properly consider the effects of subphotospheric shocks, either directly in the core of the jet or in the jet cocoon interface (JCI) \citep{gottlieb2021structure}, which is mostly where the optical photons originate in the flow. \cite{ito_mc_shocks} showed that properly simulating the shock structure and the resultant radiation can produce Band-like spectra. It is possible that these shocks could affect the optical portion of the spectra either by increasing the number of soft photons in the outflow or by upscattering optical photons to higher energies. 

Our result of observers detecting optical photons (and gamma ray photons) originating in the JCI shows the need to conduct global radiative transfer simulations which is now possible with MCRaT. Furthermore, the SRHD simulations that are analyzed should be in 3D. \cite{gottlieb2021structure} showed that SRHD simulations of GRB jets in 3D are able to properly capture the mixing between the jet core and the cocoon, which is vital to the properties of the JCI. Properly simulating this region and including it in radiative transfer calculations is important for acquiring accurate spectra and  light curves at various wavelengths. {{Furthermore, \cite{parsotan_var} showed that the photons at high density regions of the jet, which includes optical photons, are not fully decoupled from the outflow due to the relatively small domain of the SRHD simulations. As a result, future simulations that are analyzed with MCRaT should have domains large enough such that the photons can cool appropriately with the outflow \citep{parsotan_var}.}}

The improvements of properly simulating subphotospheric shocks, synchrotron emission, and being able to handle 3D SRHD simulations will be added to MCRaT in the future.

The ability to compare the photospheric model to other models in the optical region will prove fruitful in future observations of GRBs, especially as the aforementioned physical processes are added to MCRaT. The addition of  polarization observations of optical prompt emission, such as the measurements acquired by \cite{troja2017_grb160625B}, will also play an important role in ruling out potential models that describe GRB prompt emission. We explore this aspect of optical prompt emission in a companion paper \citep{Parsotan_spectropolarimetry}.

\acknowledgements 
We thank the anonymous reviewer for detailed suggestions that helped improve the content and clarity of the paper. We would also like to thank Diego L{\'o}pez-C{\'a}mara for discussions and suggestions that helped improve the paper.

TP and DL acknowledge support by NASA grants 80NSSC18K1729 (Fermi) and NNX17AK42G (ATP), Chandra grant TM9-20002X, and NSF grant AST-1907955. TP acknowledges funding from the Future Investigators in NASA Earth and Space Science and Technology (FINESST) Fellowship, NASA grant 80NSSC19K1610. Resources supporting this work were provided by the NASA High-End Computing (HEC) Program through the NASA Advanced Supercomputing (NAS) Division at Ames Research Center. Additionally, this work used the CoSINe High Performance Computing cluster,  which is supported by the College of Science at Oregon State University.  

This research has made use of the SVO Filter Profile Service (http://svo2.cab.inta-csic.es/theory/fps/) supported from the Spanish MINECO through grant AYA2017-84089. This research made use of Astropy,(http://www.astropy.org) a community-developed core Python package for Astronomy \citep{astropy:2013, astropy:2018}. \newline

\bibliography{references}
\appendix
\restartappendixnumbering
\section{The Cyclo-synchrotron Implementation}
\label{cs_appendix}
\subsection{Emitting Cyclo-synchrotron (CS) photons}
The CS emission in MCRaT is discretized into photon packets that are emitted into a given shell of fluid with properties that are taken from a given SRHD simulation. As outlined in Section \ref{cyclo-synch}, the shell corresponds to the distance traveled by the photons already in the simulation. The number of CS photons emitted in the $i^\mathrm{th}$ fluid element, with volume $dV_i$, is calculated by integrating the blackbody photon number spectrum up to the cyclotron frequency of that same fluid element, $\nu_{B',i}$. Thus, we calculate the number of emitted CS photons as
\begin{equation}
 n_{cs,i}=\int\limits_{0}^{\nu_{B',i}}\frac{u_\nu}{h \nu}d\nu\frac{dV_i}{w}=\int\limits_{0}^{\nu_{B',i}}\frac{8\pi \nu^2 }{c^3 (e^\frac{h\nu}{kT_i'}-1 )}d\nu\frac{dV_i}{w}
\end{equation}
where $w$ is the weight of the emitted photon packet, which is the number of physical photons each emitted MCRaT photon packet in the shell represents, $k$ is the Boltzmann constant, $h$ is Planck's constant, $c$ is the speed of light, and the comoving fluid temperature $T_i'=(3p_i/a)^\frac{1}{4}$, where $p_i$ is the pressure of the fluid element as acquired from the SRHD simulation and $a$ is the radiation density constant. In order to calculate this integral, we need to determine the cyclotron frequency of the fluid element which is calculated as
\begin{equation}
\nu_{B',i}=\frac{eB'_i}{2\pi m_e c}
\end{equation}
where $e$ is the electron charge, $B'_i$ is the comoving magnetic field within the cell (see Section \ref{calc_b}), and $m_e$ is the electron mass \citep{Rybiki_Lightman}. Once $ n_{cs,i}$ is calculated, we draw a random number from a Poisson distribution with a mean of $ n_{cs,i}$ to acquire the actual number of photons that are emitted into the fluid element. We get the total number of emitted photons in the shell of interest, $N_e$, by summing over the $ n_{cs,i}$ of each fluid element. If $N_e > N_{e,max}$, a user defined parameter that limits the number of emitted MCRaT photon packets, then $w$ is made larger and the aforementioned calculations are redone. If $N_e \le N_{e,max}$ then the code proceeds to assign the CS photons a random 4 momentum in the fluid rest frame, with frequency $\nu_{B',i}$, and assigns them to be at the center of their respective SRHD fluid element. 

These CS emitted photons are a pool of photons that can be upscattered as MCRaT determines which photons will be scattered. If one of these ``reserve'' photons gets chosen to be scattered, it gets replaced by an identical photon in the pool of CS photons that can be scattered. Then, we place the scattering photon randomly within its fluid element and allow it to scatter normally.

\subsection{Calculating $B'_i$}\label{calc_b}
In order to calculate $\nu_{B',i}$, we still need to determine the comoving magnetic field strength within the fluid element, $B'_i$. MCRaT provides 2 ways to calculate the magnetic field in the SRHD simulation: 1) internal energy of the flow or 2) total energy of the outflow, which the user can set. These are outlined below:
\begin{itemize}
\item Internal Energy

In the internal energy case, the comoving magnetic field energy density of the fluid element is equated to the electron thermal kinetic energy of that same element with some ratio between them, $\epsilon_{B}$.
\begin{equation}
\frac{B_i'^2}{8\pi}=\frac{3}{2}n_ikT_i'\epsilon_{B}
\end{equation}
where $n_i$ is the number density of electrons in the fluid element.

\item Total Energy

For a given plasma, the luminosity from matter and radiation can be written as $L=4\pi(\rho c^2+4p)\Gamma^2cr^2$, where c is the speed of light, $\rho$ is the plasma density, $p$ is the pressure, $\Gamma$ is the bulk Lorentz factor, and $r$ is the radius of the plasma away from its point of origin. The luminosity due to the magnetic field is $L_B=4\pi(B^2/8\pi)\Gamma^2cr^2$ where B is the magnetic field of the plasma. The fraction of energy between the magnetic outflow luminosity and the total outflow luminosity for a given fluid element, in the comoving frame, can be written as
\begin{equation}
\epsilon_{B}=\frac{L_{B,i}}{L_i}=\frac{B_i'^2}{8\pi(\rho_i c^2+4p_i)} \label{epsilon_B}
\end{equation}
If the user supplies $\epsilon_{B}$ to MCRaT then the code uses the above formula to solve for the comoving magnetic field in each fluid element.
\end{itemize}

\subsection{Rebinning Scattered CS Photons}\label{rebinning}
The MCRaT code rebins the CS photons in order to keep the number of photons that it needs to track down to a reasonable amount, which helps decrease computational time. In order to rebin the photons, MCRaT finds the photon with the smallest position in the polar direction, $\theta_\mathrm{min}$,  and the photon with the largest position in the polar direction, $\theta_\mathrm{max}$. The size of each spatial bin, $\Delta \theta$, is set by the user. The number of spatial bins, $N_\theta$, is then is determined by
\begin{equation}
N_\theta=\mathrm{floor}\bigg(\frac{\theta_\mathrm{max}-\theta_\mathrm{min}}{\Delta \theta}\bigg)
\end{equation}

The photons are also binned into energy bins, where the size of the energy bins, $\Delta E$ is determined as 
\begin{equation}
\Delta E=\frac{\log(E_\mathrm{max})-\log(E_\mathrm{min})}{N_E}
\end{equation}
where $E_\mathrm{max}$ and $E_\mathrm{min}$ are the minimum and maximum energies of the CS photons and $N_E$ is the number of energy bins as specified by the user. The number of rebinned photons is then $\leq N_\theta N_E$ since some energy and spatial bins may not contain any photons.

When rebinning the photons in polar angle and energy, the following values of photons that fall within a given energy and $\theta$ bin are averaged and then assigned to a new rebinned photon (the quantities for this rebinned photon are represented with a tilde). For a given energy bin, $\alpha$, and $\theta$ bin, $\beta$, the new rebinned photon would have the following four momentum  
\begin{equation}
\tilde{p}^\mu_{\alpha,\beta}=\frac{<E>}{c}\begin{pmatrix}
1 \\ \sin(<\theta>) \cos(<\phi>) \\ \sin(<\theta>) \sin(<\phi>) \\ \cos(<\theta>)
\end{pmatrix}
\end{equation}
where $E$ is the energy of the photon, $\theta$ and $\phi$ are the directional angles of the 4 momentum and the $<\cdots>$ denote weighted averages of the various quantities over the photons that fall within bin $(\alpha,\beta)$. The stokes parameters for the rebinned photon becomes $\tilde{s}_{\alpha,\beta}=(1,<q>,<u>,<v>)$. The new rebinned photon weight, $\tilde{w}_{\alpha,\beta}$, is calculated by summing over the photons' weights that fall within bin $(\alpha,\beta)$ and the position vector of the rebinned photon becomes
\begin{equation}
 \overrightarrow{\tilde{r}}_{\alpha,\beta}=<r>\begin{pmatrix}
  \sin(<\theta_p>) \cos(<\phi>-<\Delta \tilde{\phi_p}>) \\ \sin(<\theta_p>) \sin(<\phi>-<\Delta \tilde{\phi_p}>) \\ \cos(<\theta_p>)
 \end{pmatrix}
\end{equation}
 where $<r>$ is the average radius of the photons that lie in bin $(\alpha,\beta)$, $<\theta_p>$ is the average polar angle of the photons with the given bins and, for these same photons, $<\Delta \tilde{\phi_p}>$ is the average displacement angle taken across each photon's 4-momentum azimuthal angle, $\phi$, and its positional azimuthal angle, $\phi_p$, that is $\Delta \tilde{\phi_p}= \phi-\phi_p$. This consideration of the displacement in the positional and 4-momentum azimuthal angles is necessary to ensure that the rebinned photons are still in appropriate equilibrium with the flow and do not get erroneously upscattered. 
 
 This binning algorithm is initiated when the number of scattered CS pool photons in the MCRaT simulation exceeds some defined critical number of photons that is set by the user.
 
\subsubsection{Absorption of CS Photons}
When the time in the MCRaT simulation gets to be the time of the next frame in the SRHD simulation, MCRaT proceeds to absorb any photons that have a frequency  $\nu \le \nu_{B,i}$ of the $i^\mathrm{th}$ fluid element that they are located within. As a result, all of the CS photons that were emitted and never upscattered become absorbed and any photons that have been scattered but still satisfy  $\nu \le \nu_{B,i}$ are also absorbed. 

After MCRaT completes the absorption process, it loads the next SRHD simulation frame and re-emits the pool of CS photons based on where the remaining un-absorbed photons are located and the new fluid properties taken from the newly loaded SRHD simulation frame.

\subsection{Verifying the CS Process} \label{verify_cs}
We test our implementation of CS emission and absorption in MCRaT by conducting test simulations of radiation from a spherical outflow. As outlined in \cite{MCRaT}, we over write the values from a SRHD simulation (in memory) with those corresponding to a spherical outflow. We inject photons drawn from a  blackbody distribution at a radius corresponding to a large optical depth ($\sim 10^4$) and allow these photons to scatter and propagate through the spherical outflow, as the outflow itself moves. Simultaneously, the process of emitting and absorbing CS photons is occurring.  

\begin{figure}[]
 \centering
 \includegraphics[]{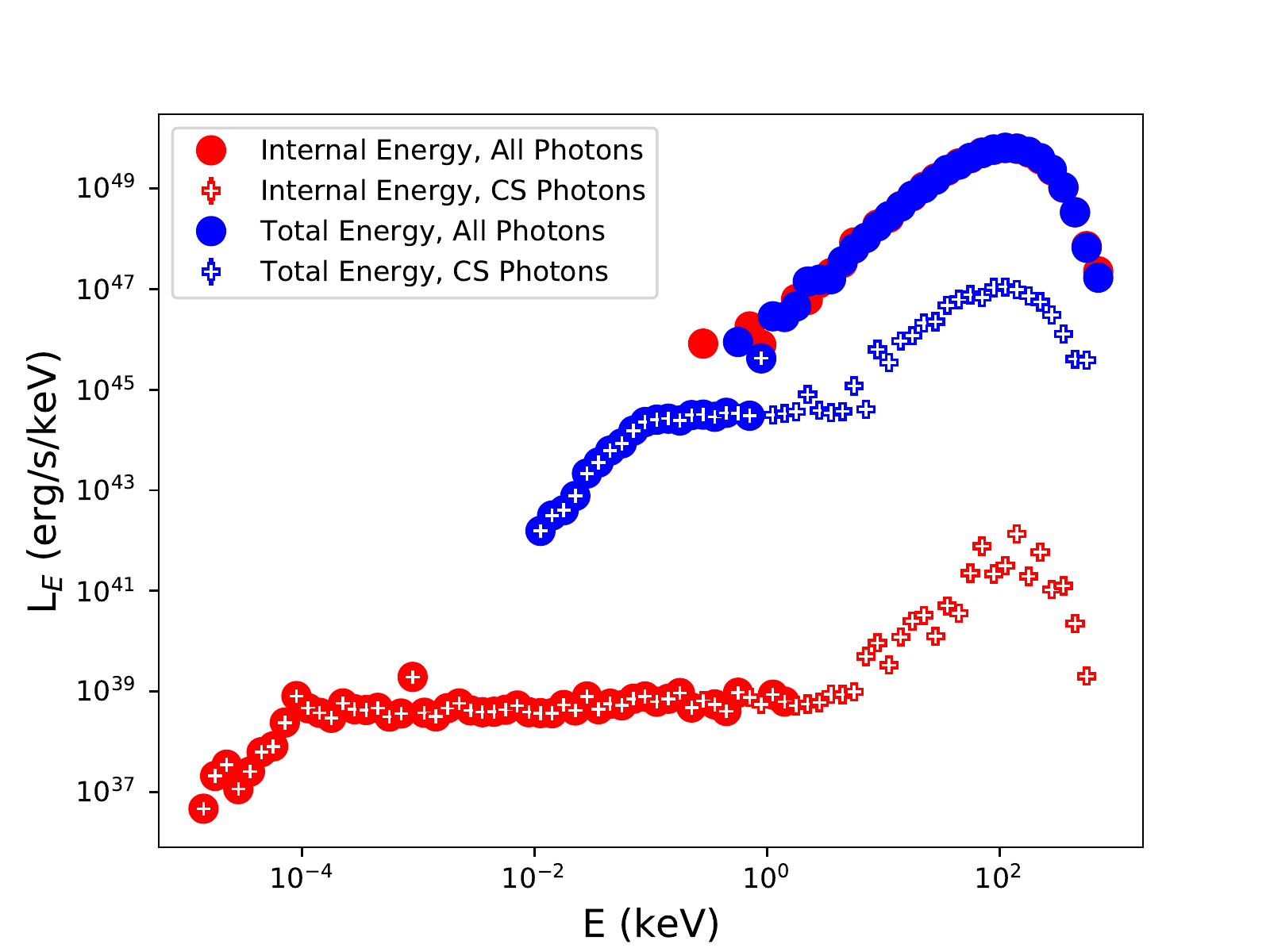}
 \caption{The two spectra produced from MCRaT simulations of spherical outflows including CS photons where the magnetic field is calculated from the total energy, shown in blue, or the internal energy of the outflow, shown in red. The circle markers show the entire spectrum consisting of the thermal photons injected as the initial condition of the MCRaT simulation and the CS photons that accumulate within the flow. The plus markers show the spectrum of just the CS photons. }
 \label{test_internal_total}
\end{figure}

In the first test, shown in Figure \ref{test_internal_total}, we look at the differences between the two methods of calculating the magnetic field. The circle markers show the entire spectrum consisting of the thermal photons injected as the initial condition of the MCRaT simulation and the CS photons that accumulate within the flow. The plus markers show the spectrum of just the CS photons. The red points are the results of the MCRaT spherical outflow simulation where the magnetic field was calculated from the internal energy, with $\epsilon_{B}=1$, while the blue points are the results from the magnetic field being calculated from the total energy of the outflow, where we set $\epsilon_{B}=0.5$. Both spectra contain $\sim 9 \times 10^4$ photons that were simulated in MCRaT. We find that the spectra consisting of solely CS photons, the plus markers, reproduces the spectrum of soft photons undergoing saturated comptonization \citep{Rybiki_Lightman}. The low energy tail has a slope of $E^{2}$ that evolves into a modified  blackbody and then a Wien spectrum at high energies\footnote{We do not simulate the emission from all regions of the spherical outflow, which produces the Wien spectrum and not the spectrum predicted by \cite{goodman1986gamma}. The \cite{goodman1986gamma} spectrum was recovered in tests of MCRaT by \cite{MCRaT} where radiation from high latitude regions of the spherical outflow was considered.}, which was also acquired by \cite{vurm2013thermalization} in their Figure 6. This outcome makes sense since these CS photons undergo $\sim 10^4$ scatterings on average, which is an indication of the optical depth in the outflow \citep{parsotan_var}. In addition to these features, we find that the initial thermal injection of photons between the test simulations stay consistent, forming a high energy Wein spectrum. The CS photon portion of the spectrum, which is produced by the internal energy method of calculating the magnetic field, lies far below the portion of the spectra formed from the thermally injected photons, as shown by the red Wien peak. In the blue spectrum, the CS produced portion of the spectrum lies slightly below the Wien portion of the spectrum. This feature shows that the total energy method of calculating the magnetic field produces more low energy photons than the internal energy method, albeit, it does not affect the Wien portion of the spectrum.

\begin{figure}[]
 \centering
 \includegraphics[]{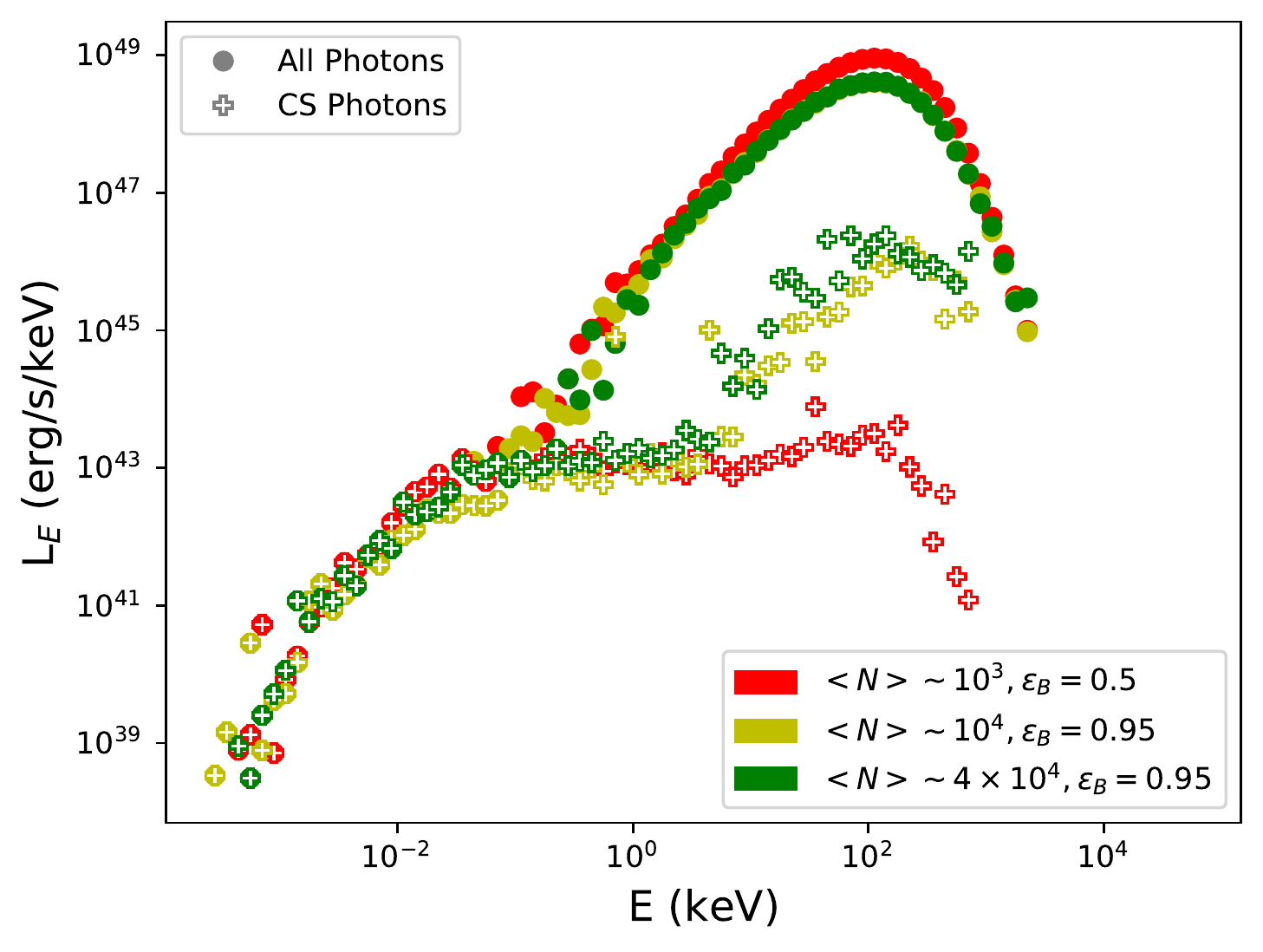}
 \caption{Spectra for 3 different test MCRaT simulations where the simulation was started at different optical depths in the outflow and each simulation used various values of $\epsilon_{B}$ (using the total energy method of calculating the magnetic field). The red points are the spectra from the simulations where the average number of scatterings that photons underwent, $<N>$, was $\sim 10^3$ and $\epsilon_{B}=0.5$, the yellow points are from the simulation where  $<N>\sim 10^4$ and $\epsilon_{B}=0.95$, and the green points are from the simulation where $<N>\sim 4 \times 10^4$ and $\epsilon_{B}=0.95$. Above $\sim 10^{1}$ keV the increase in $<N>$ affects the formation of the Wien peak, while below $10$ keV there is not much change in the spectrum. Overall, the change in $\epsilon_{B}$ has no drastic effect on the resultant spectrum due to the weak dependence of the magnetic field on this parameter.}
 \label{test_epsilonB_Nscatt_total}
\end{figure}

In our second test we use the total energy method of calculating the magnetic field and investigate the effect that changing $\epsilon_{B}$ and the optical depth in which the simulation is started have on the resultant spectrum. Figure \ref{test_epsilonB_Nscatt_total} shows the spectra for the three simulations that we ran: 1) $\epsilon_{B}=0.5$ and the simulation was started deep in the outflow where the optical depth (and thus the average number of scattering, $<N>$) $\sim 10^3$, which is shown in red, 2) $\epsilon_{B}=0.95$ and $<N>\sim 10^4$, shown in yellow, and 3) $\epsilon_{B}=0.95$ and $<N>\sim 4 \times 10^4$, shown in green. The circle markers represent the spectra constructed from the thermal injected photons and the CS photons and the plus markers represent the spectra of just the CS photons. In general, we find that as the optical depth increases, the CS photons begin to form a saturated comptonization spectrum, with the Wien peak forming between $<N>\sim 10^3-10^4$. Although the optical depth affected the formation of the Wien peak, the low energy portion of the spectra stayed relatively constant. Additionally, the low energy portion of the spectra ($\lesssim 10$ keV) did not change as the value of $\epsilon_{B}$ changed. The fact that $\epsilon_{B}$ does not have a large effect on the spectrum can be seen in Equation \ref{epsilon_B} where $B \propto \sqrt{\epsilon_{B}}$, thus an increase in $\epsilon_{B}$ by a factor of $0.95/0.5=1.9$ gives a change in the calculated magnetic field of $\sqrt{1.9}=1.37$ which does very little to change the cyclotron frequency and the number of CS photons emitted into the MCRaT simulation. 

\listofchanges
\end{document}